%% file: main.tex
\documentclass{sig-alternate-10pt}
\makeatletter
\def\ps@headings{%
\def\@oddhead{\mbox{}\scriptsize\rightmark \hfil \thepage}%
\def\@evenhead{\scriptsize\thepage \hfil \leftmark\mbox{}}%
\def\@oddfoot{}%
\def\@evenfoot{}}
\makeatother
\usepackage{graphicx,times}

\usepackage{amssymb}
\usepackage{amsmath}
\usepackage{amsbsy}

\usepackage{array}

\usepackage{balance}

\usepackage[hyphens]{url}
\usepackage{cite}

\def\squareforqed{\hbox{\rlap{$\sqcap$}$\sqcup$}}
\def\qed{\ifmmode\squareforqed\else{\unskip\nobreak\hfil
\penalty50\hskip1em\null\nobreak\hfil\squareforqed
\parfillskip=0pt\finalhyphendemerits=0\endgraf}\fi}

\newtheorem{theorem}{\bf{Theorem}}

\newtheorem{lemma}{\bf{Lemma}}
\newtheorem{definition}{\bf{Definition}}

\newcommand{\ud}{\mathrm{d}}

\begin{document}

\title{Spot Transit: Cheaper Internet Transit for Elastic Traffic}
\numberofauthors{2}
\author{
\alignauthor Hong Xu\\
\affaddr{Department of Computer Science}\\
\affaddr{City University of Hong Kong}\\
\email{henry.xu@cityu.edu.hk}
\alignauthor Baochun Li\\
\affaddr{Department of Electrical and Computer Engineering}\\
\affaddr{University of Toronto}
\email{bli@eecg.toronto.edu}
}

\maketitle

\begin{abstract}



We advocate to create a \emph{spot} Internet transit market, where transit is sold using the under-utilized backbone capacity at a lower price. The providers can improve profit by capitalizing the perishable capacity, and customers can buy transit on-demand without a minimum commitment level for elastic traffic, and as a result improve its surplus (i.e.~utility gains). We conduct a systematic study of the economical benefits of spot transit both theoretically and empirically. We propose a simple analytical framework with a general demand function, and solve the pricing problem of maximizing the expected profit, taking into account the revenue loss of regular transit when spot transit traffic hikes. We rigorously prove the price advantage of spot transit, as well as profit and surplus improvements for tier-1 ISPs and customers, respectively. Using real-world price data and traffic statistics of 6 IXPs with more than 1000 ISPs, we quantitatively evaluate spot transit and show that significant financial benefits can be achieved in both absolute and relative terms, robust to parameter values.
\end{abstract}

\input{intro}

\input{motivation}
\input{model}

\input{analysis}

\input{evaluation}
\input{discussion}
\input{related}

\input{conclusion}

\bibliographystyle{abbrv}
\bibliography{IEEEabrv,main}
\include{appendix}

\end{document}

%% file: intro.tex
\section{Introduction}
\label{sec:intro}

Internet transit has traditionally been traded with long-term contracts, where the tier-1 ISP specifies pricing and the transit customer commits a minimum level of bandwidth consumption. Two facts make this market inefficient for today's Internet. First, tier-1 ISPs typically over-provision the backbone and have a portion of the capacity under-utilized most of the time \cite{FMLC03}, which represents a bulk of the missing revenue opportunities. Second, the transit customers are increasingly unwilling to purchase transit due to sheer costs of serving the ever-increasing traffic volumes. 
Cisco estimates that busy-hour Internet traffic will increase fivefold by 2015 while average traffic will increase fourfold \cite{S11}. 

In this paper, we advocate that a \emph{spot} Internet transit market should be created, where the unused transit capacity is sold at a lower price to compliment the traditional contract-based market. 
To serve the spot traffic, the tier-1 ISP uses its under-utilized capacity that are otherwise wasted.  It provides no QoS guarantee and support for spot transit, which enables the tier-1 ISP to adopt a lower price and earn extra revenue from the available, yet perishable, bandwidth resource. It can also stop routing the spot traffic at any time, when capacity is needed for regular transit or to handle network failures. 

The transit customers, on the other hand, have the flexibility to buy transit {\em on-demand} at a discount. Spot transit is ideal for elastic traffic \cite{S95}. For example, datacenters can use it for the bulk backup and replication traffic across the Internet \cite{LSYR11,CJAZ11}. Eyeball ISPs can use it at demand valleys to support time-dependent pricing \cite{JHC11}. A lower broadband access price can then be advertised at valley periods, encouraging users to defer time-insensitive applications such as file downloading, and even out the peak demand that are particularly costly. In short, they can better cope with traffic fluctuations in a flexible and cost-efficient way.

Moreover, small ISPs that originally rely on various forms of peering or buying from transit resellers \cite{SCG11,VLFJ11} can now purchase spot transit to offload the elastic traffic, and enjoy the reachability of a tier-1 backbone with lower costs. The barrier of minimum committed data rates no longer exists. Since it does not differentiate based on the protocol or user type, spot transit is also less susceptible to network neutrality concerns spawned by some instances of paid peering \cite{level3comcast,VLFJ11}.

In this paper, we are primarily interested in the \emph{economic} aspect of the market, in particular, pricing and the resulting profit for tier-1 ISPs and consumer surplus, i.e. utility gains, for transit customers. Pricing is critical as it directly affects the incentives of both sides to participate. We take the liberty to envision that a spot market is technically feasible, and conduct a systematic study of the economical benefits of spot transit. 
The unique challenge of pricing here arises from an intriguing interplay between the spot and regular transit traffic, whose bits share the same backbone infrastructure. When the ISP tries to lower the price to attract demand, it is certainly possible that the increased spot traffic hogs up the network and causes performance degradation to the regular traffic. Therefore, the risk of \emph{demand overflow} and its impact on profit have to be explicitly taken into account. 

We solve the spot transit pricing problem for expected profit maximization of a tier-1 ISP under the classical additive random demand model \cite{PD99}, which we empirically validated using real inter-domain traffic. Further, we rigorously prove that spot transit improves both profit and consumer surplus, as long as it can be offered cheaper than regular transit. It is therefore a win-win solution for both sides of the market. We emphasize that all our theoretical results are obtained with a general demand function that captures the essential properties of any demand function. Essentially, the benefits of spot transit only depend on the characteristics of elastic traffic.

To quantitatively understand the potential of spot transit in the real world, we perform an extensive empirical evaluation based on traffic traces collected from 6 Internet eXchange Points (IXPs) in America, Europe, and Asia. The dataset of each IXP contains aggregated traffic statistics from more than 100 ISPs with peak demand between 1200~Gbps and around 200~Gbps. 
We use two canonical demand functions with real transit prices, empirical demand elasticity data, and a range of model parameters. It is demonstrated that, spot transit is typically 15\%-30\% cheaper, with more than 60\% profit improvement for tier-1 ISPs and more than 10\% surplus improvement for customers. On dollar terms, the profit and surplus gains are in the order of millions (monthly) for large IXPs and hundreds of thousands for smaller ones. The benefit is robust in the sense that a 10\% profit and surplus improvement is still observed even in the worst case.

We make three original contributions. First, we propose spot transit, a transit market that allows customers to buy underutilized transit capacity on-demand at a discounted price (Sec.~\ref{sec:motivation}). Second, we propose a simple and practical analytical framework that strikes a balance between economic theory and realistic aspects of Internet transit. With a general demand function (Sec.~\ref{sec:model}), we characterize the optimal price as a function of the provision cost, overflow penalty, demand elasticity, and the predictability of traffic. We theoretically establish the price advantage and efficiency gains of spot transit (Sec.~\ref{sec:analysis}). Third, we empirically evaluate spot transit using real traffic statistics and price data. We validate our demand model, and obtain realistic parameters for demand functions by fitting empirical traffic data into canonical economic models, and compellingly demonstrate the significant benefits of spot transit (Sec.~\ref{sec:evaluation}). Towards the end, we discuss practical issues towards implementing such a market (Sec.~\ref{sec:discussion}).

%% file: motivation.tex
\section{Background and Motivation}
\label{sec:motivation}

The Internet backbone consists of a small number of tier-1 ISPs, each owning a portion of the global infrastructure, and can reach the entire Internet solely via settlement-free interconnection, i.e. {\em peering}. They provide transit services to {\em transit customers} for monetary returns. It is widely known that the tier-1 backbone is largely under-utilized (under 50\%) due to over-provisioning \cite{FMLC03}. Over recent years, though traffic has been growing at a 40\%--50\% annual rate \cite{LIMO10,S11}, backbone capacity has been increasing worldwide at an equivalent pace according to \cite{T11}. As a result, underutilization continues to exist, with average and peak link utilization on major backbone lines stay virtually unchanged \cite{T08}. 

The under-utilized backbone capacity represents a bulk of missing revenue opportunities for tier-1 ISPs. Network capacity is inherently a {\em perishable} resource: bandwidth that is not used is lost forever. On the other hand, despite the constant decline of transit prices \cite{T11}, customers face enormous transit costs due to the rapid-growing traffic. New forms of peering \cite{VLFJ11} and smart traffic engineering schemes \cite{LSYR11,XL12} are continuously being developed to cut the transit bill. This demonstrates the inefficiency of the current market, and calls for novel solutions that offer better ways of trading transit.

Motivated by these observations, we propose to create a {\em spot} transit market, where the unused capacity is sold at a discount without SLAs regarding network availability, route stability, etc. Transit customers can purchase spot transit on-demand without entering a contract first. The spot transit market compliments the regular contract based market. It suits well to serve the elastic traffic, such as the bulk replication and backup traffic across datacenters and most residential broadband traffic, because it can tolerate delay and loss and is much more price sensitive \cite{VLFJ11}.

\begin{figure}[hbtp]
    \centering
    \includegraphics[width=1\linewidth]{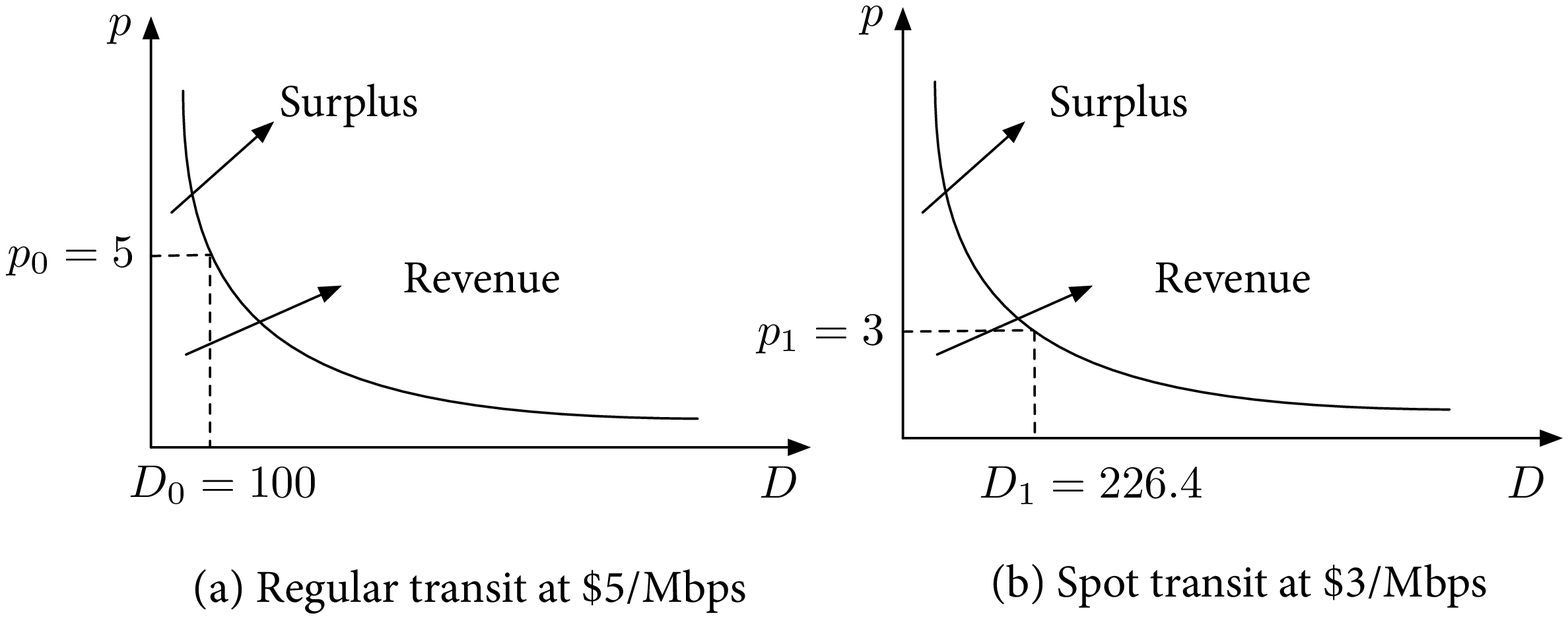}
    \caption{The benefits of the spot transit market. It improves the social welfare by improving the revenue for tier-1 ISPs, and surplus for transit customers. The demand curve shown is $D = 1313.26 \cdot p^{-1.6}$.}
    \label{fig:benefits}
\end{figure}

To intuitively see the potential of spot transit market, Figure~\ref{fig:benefits} shows a typical iso-elastic demand curve \cite{MW00,VLFJ11} for a tier-1 ISP's elastic traffic. We choose the demand function such that at a regular transit price of \$5/Mbps, the elastic traffic is 100~Gbps. The revenue is \$500,000 per month. If the demand were to be served by the spot transit market, since the tier-1 ISP does not need to provide QoS guarantees, it can afford a lower price of, say, \$3/Mbps, with a 40\% discount. At such a low price, demand rises to around 226.4~Gbps, and our tier-1 ISP collects around \$680,000. Thus, by creating the spot transit market and optimizing price, the tier-1 ISP attains a 36\% profit increase. This represents a strong financial incentive. 

The spot transit market also benefits transit customers in terms of consumer surplus, which is the difference between the total amount that they are willing to pay and the actual amount that they do pay at the market price \cite{MWG95}. As depicted in Figure~\ref{fig:benefits}, consumer surplus is greatly improved with a lower transit price. 

Therefore, the spot market achieves higher efficiency and improves social welfare. The underlying reason is that in the conventional transit market, elastic traffic is priced together with inelastic traffic that has higher costs to serve. By pricing the elastic traffic separately with spot transit that has no SLAs, tier-1 ISPs are able to adopt a lower price, which in turns attracts even more demand and increases profit. 

The demand increase with spot transit can be explained by at least two factors: competition with peering and transit reselling. The low price and on-demand feature of spot transit can make it more appealing than peering or paid peering, considering the performance benefits and network reachability provided by the tier-1 backbone. Further, small ISPs usually find it difficult to purchase transit directly from tier-1 ISPs due to the minimum committed data rate requirement, and rely on transit resellers instead. With spot transit, tier-1 ISPs are able to collect additional profits from these small ISPs by bypassing the transit resellers in the middle. In all, we expect that spot transit compliments the traditional transit business of a tier-1 ISP.

%% file: model.tex
\section{Model}
\label{sec:model}

In this section, we introduce the theoretical model for our analysis of the optimal pricing, and the benefits of spot transit.

\subsection{Spot transit market}
\label{subsec:spot_market}

We consider a spot transit market with multiple tier-1 ISPs. This corresponds to an oligopoly scenario. Numerous game models can be adopted \cite{MWG95} and each considers specific competition scenarios. Our objective is to study spot transit under a general model that captures the key aspects of the market. For the applicability of results and ease of exposition, we choose to focus on a representative tier-1 ISP, with the {\em residual} demand that is not met by other competitors in the market. Residual demand is a common concept in microeconomics in studying the pricing problem for a firm operating in a competitive market \cite{VLFJ11,MW00}. 

Though residual demand does not model the full dynamic interactions between ISPs, it accounts for the availability of substitutes and switching costs. It also allows a faithful empirical verification, since the real-world price and demand data collected reflect the effect of competition. The same approach is also adopted in \cite{VLFJ11}. 

\subsection{Demand}
\label{subsec:demand_model}
One of the challenges of conducting economic analysis in networking is the lack of a good model for the residual demand, or simply demand, in general. Not much public information is known about the Internet transit market in particular for business reasons. Here we blend theory with practice, and use classical demand models from economics with empirical justifications based on real traffic data. We believe such an approach not only makes the analysis tractable, but also the results practically relevant.

\subsubsection{A model for billable demand}
\label{subsubsec:95-demand}

\begin{figure}[tbp]
    \centering
    \begin{minipage}[t]{1\linewidth}
        \includegraphics[width=1\linewidth]{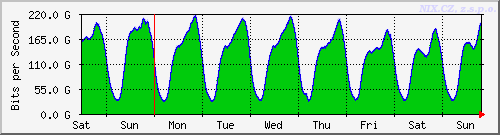}
    \end{minipage}
    \begin{minipage}[t]{1\linewidth}
        \includegraphics[width=1\linewidth]{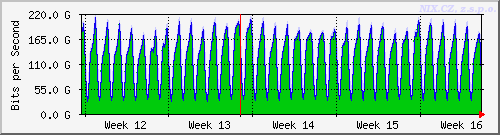}
    \end{minipage}
  \caption{Weekly and monthly aggregated traffic at NIX, an Internet exchange in Czech \cite{N12}.} 
    \label{fig:showcase}
\end{figure}

We define spot transit demand $D$ as the actual \emph{billable} amount of bandwidth, i.e.~the 95-percentile demand. Other pricing schemes, such as volume pricing, can be studied in a similar way. The aggregated traffic of a tier-1 ISP often exhibits a diurnal pattern with high predictability \cite{FMLC03,PTZD03}. For example Figure~\ref{fig:showcase} plots the aggregated traffic at the Neutral Internet eXchange (NIX) in Czech Republic \cite{N12}. The diurnal pattern is clearly visible in both weekly and monthly scales. Naturally, one can thus employ statistical methods over traffic time series to estimate the billable demand, with a small error that arises from the inherent randomness of demand.

Inspired by this observation, we adopt the classical approach in economics \cite{PD99} to model the uncertainty of billable demand in an additive fashion. Specifically,
\begin{equation}\label{eqn:95-demand}
  D(p,\epsilon) = d(p) + \epsilon,
\end{equation}
following \cite{M59}. Here, $p$ is the spot transit price in \$/Mbps, $d(p)$ is the price-dependent demand function that models the billable demand (more details in Sec.~\ref{subsubsec:d_p}), and $\epsilon$ is a random variable defined over $[A,B]$ to model the inherent demand uncertainty. Thus randomness in demand is price independent. That is, the shape of the demand curve is independent of the price, while the mean and variance of the demand distribution are affected by the price. We provide empirical justifications for our demand model using real-world traffic data in Sec.~\ref{sec:model_validation}. 

\subsubsection{Demand function $d(p)$ and elasticity}
\label{subsubsec:d_p}
Many demand functions $d(p)$ are valid in economic analysis. Instead of choosing some specific functions to work with, our analysis assumes a general demand function. We only require that $d(p)$ is a continuous, twice differentiable, decreasing, and convex function, i.e. $d'(p)<0$, $d''(p)\ge 0$. Monotonicity and convexity are general characteristics of the demand-price interaction. These assumptions are quite reasonable and commonly accepted in the literature.

A useful concept related to demand is its {\em elasticity}. Elasticity measures the responsiveness of demand to a change in price, and is defined at a price point $p$ as
\begin{equation}\label{eqn:elasticity}
    \sigma (p) = - \frac{p\cdot d'(p)}{d(p)}.
\end{equation}
We can observe that 
\begin{equation}\label{eqn:sigma'}
    \sigma'(p) \ge 0
\end{equation}
since $d'(p)<0$ and $d''(p)\ge 0$, i.e.~elasticity is non-decreasing in $p$.

Next we show two canonical demand functions that satisfy our assumptions.

{\bf Iso-elastic demand.} The iso-elastic demand, or constant elasticity demand, is a well-known demand function derived from the {\em alpha-fair} utility function \cite{MW00,VLFJ11}, which is often used to model Internet user activity. As the name suggests, elasticity is constant for every price point. 
\begin{equation}\label{eqn:ced}
  d(p) = v\cdot p^{-\alpha}, v>0,\alpha>1.
\end{equation}
$v$ can be interpreted as the {base demand} that controls the magnitude of demand. $\sigma(p)=\alpha$, where $\alpha$ denotes the constant elasticity: a higher value represents higher elasticity. As discussed above, demand here is the residual demand, and high elasticity can also indicate that the market is more competitive, and substitutes are readily available. 

{\bf Linear demand.} The simple linear function of demand is also popular \cite{MWG95}:
\begin{equation}\label{eqn:lineard}
  d(p)=v-\alpha p, v>0, \alpha >0.
\end{equation}
Here $v$ is the base demand. Elasticity of linear demand is $\sigma(p) = \frac{\alpha p}{v-\alpha p}$. In contrast to iso-elastic demand, now elasticity decreases in price. Thus it captures the phenomena that sometimes, demand is less sensitive to a price change when the price is already low (recall \eqref{eqn:sigma'}).

\subsection{Surplus, profit, and social welfare}
\label{subsec:profit_model}

To comprehensively study the economical benefits of spot transit we consider three metrics, namely {\em consumer surplus}, {\em provider profit}, and {\em social welfare}. As we have seen from Figure~\ref{fig:benefits}, consumer surplus, or simply surplus is the utility gain obtained by customers due to the purchase of Internet transit. Specifically,
\begin{equation}\label{eqn:surplus}
    S(p) = \int_p^{\infty} (x-p) d(x)\ud x.
\end{equation}
In other words, the surplus equals the amount customers are willing to pay, minus the actual cost of purchase at $p$. It is evident that $S(p)$ increases as $p$ decreases, i.e.~a price reduction is always beneficial for customers.

Next we characterize provider profit as a function of the spot transit price $p$. We consider the scenario where the tier-1 ISP allocates a fixed portion of the unused backbone bandwidth to offer spot transit services. This amount is defined as the {\em capacity} of spot transit $C$, and can be safely used without affecting the ISP's regular business. A proper choice of $C$ can be determined by the ISP profiling its network utilization. 

The notion of {\em capacity} here is {\em not} a rigid resource constraint. Since demand is inherently random and both the spot and regular transit traffic share the same backbone, nothing prevents the spot traffic from breaking through the capacity $C$ and using the capacity reserved for regular transit. Such a \emph{demand overflow} scenario may negatively affect the regular transit traffic and thus the ISP's revenue. To model the revenue loss, a penalty of $m>0$ in \$/Mbps is incurred whenever demand exceeds the capacity, i.e. $D(p,\epsilon)>C$. An equivalent interpretation is to treat it as modeling the surplus loss of the regular transit customers due to overflow. $m>p^C$, where $p^C$ denotes the price at which $d(p)=C$. Thus, the penalty is large enough so that at the optimal operating point, the expected demand is smaller than $C$.

We let $f(\cdot)$ denote the probability density function of the demand uncertainty $\epsilon$ defined over $[A, B]$. In practice $\epsilon$ can often be approximated by a Gaussian random variable. To make sure that positive demand is possible, we require $B<C$ which holds naturally since the randomness is small in magnitude compared to capacity. These assumptions will also be verified using real traffic traces in Sec.~\ref{sec:model_validation}. 

We can now formally define the spot transit profit. If demand does not exceed the capacity, then profit is simply $(p-r)D(p,\epsilon)$, where $r>0$ is the unit cost of spot transit. Otherwise the profit consists of a positive component from serving the spot transit and a negative component representing the demand overflow loss, and is written as $(p-r)D(p,\epsilon) - m(D(p,\epsilon)-C)$. The profit function, $R(p,\epsilon)$ can then be expressed as
\begin{equation*}
  R(p,\epsilon)=\left\{ \begin{array}{cl}
                      (p-r)(d(p)+\epsilon), & \epsilon \le C-d(p)\\
                      (p-r-m)(d(p)+\epsilon) + mC, & \epsilon > C-d(p)
                      \end{array}\right.
\end{equation*}
 
The expected profit is:
\begin{multline}\label{eqn:profit}
  E[R(p)] = \\
   (p-r)d(p) - m\int_{C-d(p)}^B (d(p)-C+u)f(u)\mathrm{d}u
\end{multline}
Define $\Phi(p)=(p-r)d(p)$, and $\Lambda(p)=m\int_{C-d(p)}^B (d(p)-C+u)f(u)\mathrm{d}u$. $\Phi(p)$ represents the \emph{risk-free} profit when demand is deterministic. $\Lambda(p)$ is the loss function at an average cost of $m$ when demand overflow happens. The overall expected profit is the difference between the two.
\begin{equation}\label{eqn:profit_short}
  E[R(p)]=\Phi(p) - \Lambda(p),
\end{equation}

Finally, the {\em social welfare} of a market is defined as the sum of consumer surplus and provider profit. For the spot transit market, its social welfare $\Psi(p)$ is
\begin{equation}\label{eqn:welfare_spot}
    \Psi(p) = S(p) + E[R(p)].
\end{equation}

%% file: analysis.tex
\section{An Economic Analysis}
\label{sec:analysis}

In this section, we present our analysis on the optimal pricing of spot transit for profit maximization, as well as the resulting profit, surplus and social welfare, to theoretically demonstrate its benefits.

\subsection{Pricing for profit maximization}
\label{sec:pricing}

With spot transit, the very first question we need to answer is, how do we price it? Our ISP needs to determine a price to maximize its expected profit with the presence of demand uncertainty, taking into account the risk of demand overflow and its monetary impact. The profit maximization problem can then be formulated: 
\begin{gather}\label{opt:monopoly}
  \max_{p}\ E[R(p)] \\
   \text{s.t. } \eqref{eqn:profit} \nonumber 
\end{gather}

For efficient price determination, the optimization problem must have an efficient solution algorithm. The most useful criterion for this property is convexity: minimizing a convex function, or equivalently maximizing a concave function over a convex constraint set. However, we show that this condition is not satisfied for our problem.

Consider the first-order derivative of $E[R(p)]$, which can be obtained by applying the Leibniz integral rule:
\begin{equation}\label{eqn:E'}
	E'[R(p)]  = d(p) + d'(p)\left(p-r-m\cdot\Pr\big(\epsilon>C-d(p)\big)\right).
\end{equation}
It can be observed that the term in $(\cdot)$ is positive and decreasing in $p$, and thus the term $-(\cdot)d'(p)$ is decreasing in $p$. However, $d(p)+p\cdot d'(p)$ is not monotonically increasing or decreasing. Therefore, $E[R(p)]$ is neither convex nor concave in $p$.

Luckily, we can still prove that the first-order optimality condition $E'[R(p)] = 0$ is a necessary and sufficient condition for the optimization \eqref{opt:monopoly}, with a very mild assumption of {\em quasiconcavity}. 
\begin{definition}\label{def:quasiconcavity}
    A function $g(x)$ is (strict) quasiconcave if and only $g'(x)(x'-x) > 0$ whenever $g(x')>g(x)$ (p.934, \cite{MWG95}).
\end{definition}

That is, a quasiconcave function is either decreasing, increasing, or there exists $x^*$ such that $g$ is decreasing for $x<x^*$ and increasing for $x>x^*$. Thus, quasiconcavity is a generalization and relaxation of concavity. If a function is not monotone, quasiconcavity guarantees that it has a unique global maximum. In other words, it alleviates the burden of considering the second-order condition by ensuring that the sufficient first-order condition is necessary even without the strong concavity assumption \cite{MWG95,AE61}.

\begin{lemma}\label{lem:quasiconcavity}
    $E[R(p)]$ as in \eqref{eqn:profit} is quasiconcave with the general demand function $d(p)$.
\end{lemma}
The proof can be found in Appendix~\ref{app:lem:quasi}. 

Therefore, we can efficiently solve the optimal pricing problem by setting the first-order derivative of $E[R(p)]$ to zero. 
\begin{theorem}\label{thm:monopoly}
    The optimal price price $p^*$ of the profit maximization problem \eqref{opt:monopoly} is determined uniquely as the solution to the first-order condition, i.e.
    \begin{equation}\label{eqn:p*}
         p^* = r + m\cdot \Pr \big(\epsilon > C-d(p^*)\big)  - \frac{d(p^*)}{d'(p^*)}.
    \end{equation}
\end{theorem}

For example, the optimal price of iso-elastic demand \eqref{eqn:ced} satisfies 
\begin{equation}\label{eqn:p*_ced}
  p^*= \frac{\alpha}{\alpha-1}\left( r + m\cdot\Pr\big(\epsilon > C-v(p^*)^{-\alpha} \big) \right).
\end{equation}
The optimal price of linear demand \eqref{eqn:lineard} satisfies
\begin{equation}\label{eqn:p*_linear}
    2 p^* = r + m\cdot\Pr\big(\epsilon > C-v+\alpha p^*\big)  + \frac{v}{\alpha} .
\end{equation}

Several interesting and economically satisfying observations can be made from Theorem~\ref{thm:monopoly}. 
First, the optimal price $p^*$ increases with the provision cost $r$, which is straightforward. Second, $p^*$ also increases with the overflow penalty $m$. Third, $p^*$ increases with $-\frac{d(p^*)}{d'(p^*)}$, which equals $\frac{p^*}{\sigma(p^*)}$ from \eqref{eqn:elasticity}. That is, the ISP can set a high price if demand elasticity is low due to weak market competition or the uniqueness of its service, and vice versa. Finally, we can see that $p^*>p'$, where $p'$ is the profit maximizing price without demand uncertainty ($m=0$). It shows that with demand uncertainty, a tier-1 ISP needs to charge a higher price in order to cover the damage of demand overflow.  

Having solved the pricing problem, we now would like to study the price advantage of spot transit, i.e. whether, or when it can be offered at a cheaper price than regular transit. To address this question, we need to first understand the regular transit price, which is set prior to the introduction of spot transit. For a rational tier-1 ISP, we can interpret $\bar{p}$ as the profit maximizing price for the aggregated traffic demand $\bar{d}(p)$ when it is solely served by the regular transit, which amounts to the following:  
\begin{gather}\label{eqn:total_profit}
    \bar{p} = \arg\max\ (p-\bar{r})\bar{d}(p),
\end{gather}
where $\bar{r}$ is the provision cost of regular transit. Notice that although regular and spot transit share the same infrastructure cost, $\bar{r} > r$ because spot transit does not have any QoS support or financial overhead of contract negotiation. Since $\bar{d}(p)$ includes both inelastic and elastic traffic demand, its elasticity is smaller, i.e. $\bar{\sigma}(p)<\sigma(p)$ for any $p$. Then we can prove the following.
\begin{theorem}\label{thm:p*_properties}
    $p^*<\bar{p}$, i.e. the spot transit price is less than the regular transit price, if the following is satisfied:
	\begin{equation}\label{eqn:price_condition}
		r \le \bar{r} - m\left(1-\sigma(p^*)^{-1}\right)\frac{\theta^2}{\theta^2+(C-d(p^*)-\mu)^2},
	\end{equation}
	where $\mu$ and $\theta$ are the mean and standard deviation of $\epsilon$, respectively.
\end{theorem}

The proof is in Appendix~\ref{app:thm:p*}. Theorem~\ref{thm:p*_properties} confirms the intuition that spot transit is cheaper than regular transit as long as the cost difference between them is large enough. The condition \eqref{eqn:price_condition} is sufficient but not necessary. It is easy to satisfy, since $r < \bar{r}$ always holds and $\left(1-\sigma(p^*)^{-1}\right)$\footnote{$\sigma(p^*)>1$ because price reduction is only possible to generate positive gains if demand increases at least proportionally in response.} is small in practice. The term $\frac{\theta^2}{\theta^2+(C-d(p^*)-\mu)^2}$ bounds the tail probability $\Pr(\epsilon>C-d(p^*))$, and is also small given the demand randomness $\epsilon$, i.e. its standard deviation $\theta$ is small. Therefore, theoretically, spot transit can be offered at a discount in most of the cases with a general demand function. 

We wish to emphasize that this result depends only on two defining characteristics of elastic traffic that spot transit serves, i.e. low cost \eqref{eqn:price_condition} and high elasticity $\sigma(p)>\bar{\sigma}(p)$. As will be shown soon in Sec.~\ref{sec:efficiency}, they also guarantee the economical efficiency gains of spot transit. This demonstrates the generality of our results that does not depend on the specific forms the demand function may take.

\subsection{Surplus, profit, and social welfare}
\label{sec:efficiency}

Now we turn to analyzing the efficiency of the spot transit market. We seek to answer the question: at the optimal price $p^*$, can spot transit improve surplus, profit, and eventually social welfare? 

Without loss of generality, we assume that $p^* < \bar{p}$ holds. First, as discussed in Sec.~\ref{subsec:profit_model}, surplus increases when price decreases, and the following follows from Theorem~\ref{thm:p*_properties}.
\begin{lemma}\label{lem:surplus}
     $S(p^*) > S(\bar{p})$.
\end{lemma}
That is, spot transit improves surplus for elastic traffic given its price advantage. Thus it is beneficial for customers.

Next from the provider's perspective, we wish to know whether spot transit is more attractive than regular transit for elastic traffic. That is, whether the optimal profit is larger than that using regular transit. Without spot transit, the expected profit that could be collected from elastic traffic at the regular transit price $\bar{p}$ is: 
\begin{equation}\label{eqn:benchmark_profit}
    E[{R}(\bar{p})] = (\bar{p} - \bar{r})  d(\bar{p}),
\end{equation} 
where $d(\bar{p})$ is the demand that would occur at $\bar{p}$. 

\begin{lemma}\label{lem:profit}
    $E[R(p^*)] > E[{R}(\bar{p})]$.
\end{lemma}
The proof is in Appendix~\ref{app:lem:profit}. This lemma confirms that the spot transit market is not only beneficial for customers, but also profitable for ISPs, as long as it can be offered at a lower price than the regular transit. The reason of the profitability is that since elastic traffic is more price sensitive, a price reduction can potentially attract more demand and the end result is a net profit increase despite the negative impact of demand overflow on regular transit.

Combining Lemma~\ref{lem:surplus} and \ref{lem:profit} we have
\begin{theorem}\label{thm:welfare}
    $\Phi(p^*) > \Phi(\bar{p})$. Spot transit improves social welfare by improving consumer surplus and ISP profit. 
\end{theorem}

Therefore, we have theoretically proved that spot transit is more efficient than the conventional market for elastic traffic, taking into account demand uncertainty and the overflow loss. Both ISPs and transit customers have clear incentives to participate in this new market. Given the flexibility in purchasing, spot transit represents an economically viable and attractive market solution for Internet transit, especially for elastic traffic. Our analysis is valid for general demand functions, and can be expected to hold in most realistic cases.

%% file: evaluation.tex
\section{Evaluation }
\label{sec:evaluation}

\begin{table*}
  \small
  \centering
    \begin{tabular}{ | c | c | c | c | c | c |c|c|}
    \hline 
    IXP & Acronym & Country & \# of members & Peak (Gbps) & Average (Gbps) & Error mean $\mu$ & Error s.d. $\theta$\\ \hline
        London IX \cite{L12} & LINX (LN) & U.K. & 407 & 1200 & 797.1 & -15.9278 & 174.8157\\
        Moscow IX \cite{M12} & MSKIX (MSK) & Russia & 353 & 688.5 & 416 & 2.2313 & 115.0810\\
        Neutral IX (Prague) \cite{N12} & NIX (N) & Czech & 54 & 217.8 & 129.6 & -1.2458 & 30.2338\\
 New York International IX \cite{N12a} & NYIIX (NY) & U.S. & 128 & 205.9 & 157.7 & 3.9486 & 26.0743\\
                   Spain IX \cite{S12} & ESPANIX (ES) & Spain & 58 & 198 & 172.5 & -1.0476 & 22.3824\\
               Hong Kong IX \cite{H12} & HKIX (HK) & China & 168 & 180 & 119.8 & 1.7689 & 22.0919\\
    \hline  
    \end{tabular}
    \caption{IXPs studied and statistics of their traffic and week-ahead prediction errors.}
\label{table:ixps}
\end{table*}

In this section, we present an empirical evaluation of spot transit based on real-world operational traffic information we collected from 6 representative Internet eXchange Points (IXPs) located in America, Europe, and Asia. One significant hurdle of network economic analysis is the lack of empirical data. ISPs, especially tier-1 ISPs, are reluctant to share their traffic data publicly. A few European ISPs serving academic institutions do publish their traffic statistics online. Their size, traffic volume, and characteristics, however, do not faithfully reflect those of the tier-1 ISPs.

We use the traffic statistics of large IXPs to represent the tier-1 ISP traffic. IXPs are physical infrastructure facilities through which ISPs peer with each other. They are located in key hub locations around the world, and serve a significant portion of the Internet traffic \cite{LIMO10}. A large IXP typically has over one hundred member ISPs, and interconnecting hundreds of Gb/s commercial, academic, and residential traffic. Thus the scale and traffic characteristics are similar to those of a tier-1 ISP. 

Though IXPs solely serve peering traffic, empirically it has been verified that peering and transit traffic share similar temporal patterns with respect to peak times, peak-to-valley ratios, etc.~\cite{SCG11}. Essentially, both are driven by the same end-user behavior, and it is intuitive that they are statistically similar. Thus, we believe the first-order estimation of transit traffic using peering traffic of IXPs is appropriate as a starting point of our performance evaluation. In the following, we use the IXP traffic to represent the regular transit demand $\bar{d}(\bar{p})$ at the regular transit price for tier-1 ISPs.

\subsection{Dataset description}
\label{sec:dataset}

IXP data is more accessible since many publish their aggregate traffic statistics online. Usually the incoming and outgoing traffic time series are reported and updated every 5 minutes. 
We manually inspect the webpages of large IXPs \cite{IXP}, and handpick 6 representative ones across the globe that publish traffic statistics using the standard \texttt{mrtg/rrdtool} visualization tool \cite{mrtg} with a reasonable time granularity\footnote{Some large IXPs, such as Deutscher IX, Amsterdam IX, and JPNAP only publish daily and/or yearly traffic stats that are too coarse to analyze.}. We crawl the websites of these IXPs to collect the weekly aggregated traffic statistics images. All the data is collected in 2012 and is more recent than the dataset in \cite{SCG11}. Table~\ref{table:ixps} lists all the IXPs we studied in this paper.

Traffic data is published as {\tt png} images using {\tt mrtg}, and is not readily available in numerical forms. We follow the approach of \cite{SCG11} to use an optical character recognition (OCR) program to read the {\tt png} images and output the numeric array containing the traffic time series. Each IXP uses a slightly different {\tt mrtg} configuration, including the size, bit depth, and color representation. Thus, we modified the software provided by \cite{SCG11} to handle each IXP's {\tt png} image individually. The raw image files, the numeric data, and the software for converting {\tt png} images are available in \cite{ixpdata}. The basic traffic information is shown in Table~\ref{table:ixps}.

\subsection{Demand model validation}
\label{sec:model_validation}
As a first step, we conduct an empirical validation of our demand model stated in Sec.~\ref{subsubsec:95-demand}.
Recall from \eqref{eqn:95-demand} that we model the 95-percentile demand as the sum of the demand function $d(p)$ and a random variable $\epsilon$ to model the uncertainty. Since the aggregated traffic has a clear diurnal pattern, an intuitive justification of this model can be provided if the ISP can accurately estimate its 95-th percentile demand based on the traffic time series, with a small error term to account for the unpredictable dynamics that corresponds to $\epsilon$ \cite{B06,PTZD03}. 

We assume that the tier-1 ISP of interest uses the most recent history to estimate/predict the future demand time series, the simplest regression method \cite{B06}. Once the entire time series can be predicted, its 95-percentile can be readily obtained. More complex algorithms can yield more accurate prediction for a longer time window \cite{PTZD03}, which is beyond the focus of this paper. Thus, if the prediction window size is $T$, the future demand at time $t$ is $D_t = D_{t-T}$.

\begin{figure}[tbp]
    \begin{minipage}[t]{0.49\linewidth}
    \centering
    \includegraphics[width=1\linewidth]{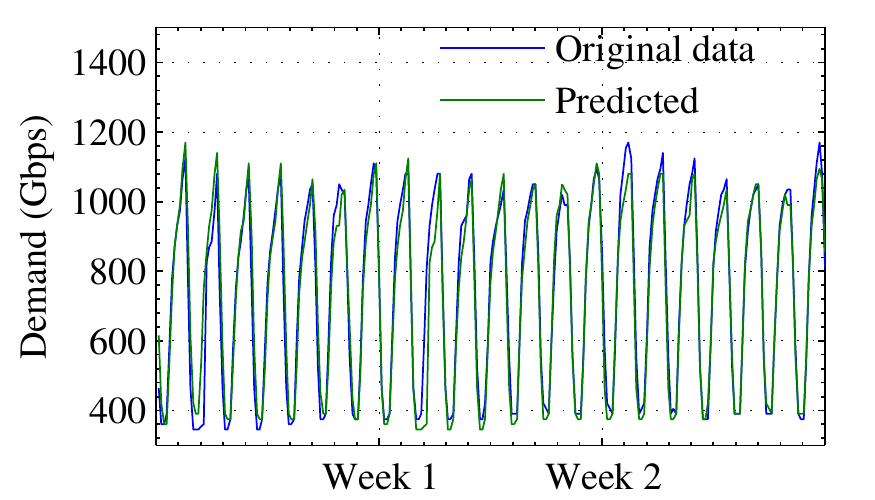}
\vspace{-3.3mm}
    \caption{Week-ahead prediction of the LINX traffic.}
    \label{fig:linx-week}
    \end{minipage}
    \begin{minipage}[t]{0.49\linewidth}
    \centering
    \includegraphics[width=1\linewidth]{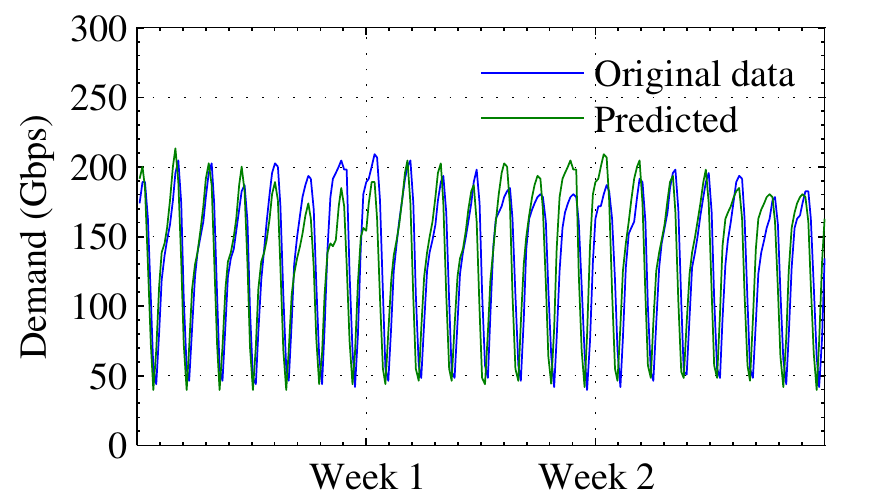}
\vspace{-3.3mm}
    \caption{Week-ahead prediction of the NIX traffic.}
    \label{fig:nix-week}
    \end{minipage}
\vspace{-3.3mm}
\end{figure}

\begin{figure}[tbp]
    \begin{minipage}[t]{0.49\linewidth}
    \centering
    \includegraphics[width=1\linewidth]{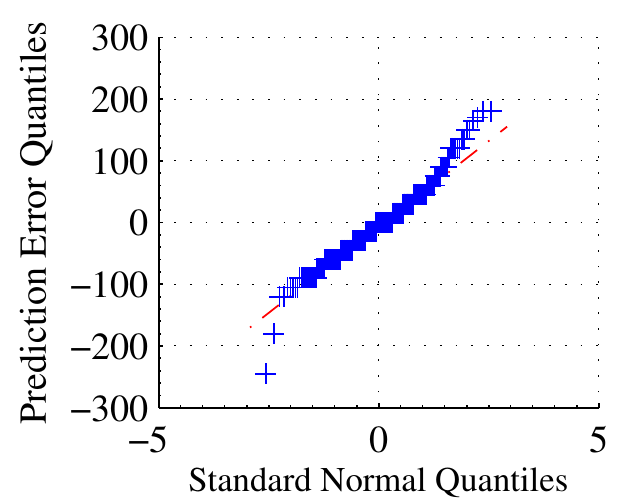}
    \caption{Q-Q plot of week-ahead prediction error on the LINX trace.}
    \label{fig:linx-week-error}
    \end{minipage}
    \begin{minipage}[t]{0.49\linewidth}
    \centering
    \includegraphics[width=1\linewidth]{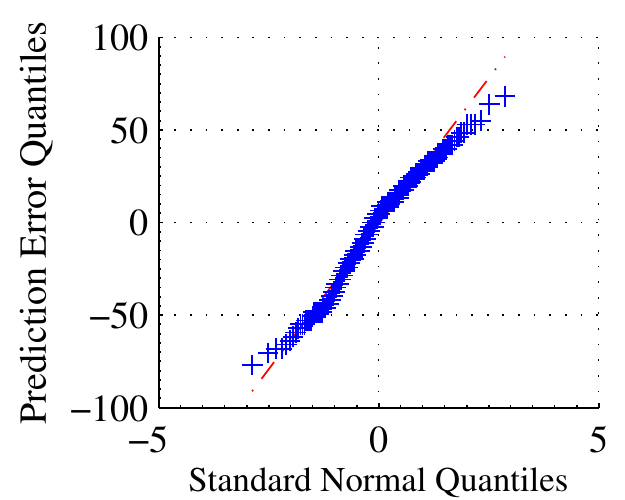}
    \caption{Q-Q plot of week-ahead prediction error on the NIX trace.}
    \label{fig:nix-week-error}
    \end{minipage}
  \end{figure}

We run this week-ahead prediction on all 6 IXPs. Figure~\ref{fig:linx-week} and \ref{fig:nix-week} show the week-ahead ($T=7$ days) prediction result of LINX and NIX traffic for an example. 
We can observe that simple week-ahead prediction based on the most recent history is fairly accurate. Figure~\ref{fig:linx-week-error} and \ref{fig:nix-week-error} show the Q-Q plots of the prediction errors. They lie closely on a linear line, suggesting that the error term ${\epsilon}$ behaves much like a Gaussian random variable. 

The error mean $\mu$ and standard deviation $\theta$ for the week-ahead prediction on all 6 IXPs are shown in Table~\ref{table:ixps}. The mean is close to zero and the standard deviation is small compared to the predicted data. Since the demand series can be accurately predicted with a small error, the 95-percentile demand can be readily calculated with the same error statistics. This validates our demand model.

Note that although the IXP data contains both elastic and inelastic traffic, essentially our demand model is based upon the observation that demand of a tier-1 ISP is highly predictable due to multiplexing, which is valid for both traffic types. We use the 95-percentile traffic from the IXP data to represent the aggregated regular transit demand $\bar{d}(\bar{p})$ at the regular price $\bar{p}$. We adjust $\beta\in[0.2,0.7]$ to obtain elastic traffic out of the aggregated, where $\beta$ denotes the relative proportion of elastic traffic. $\mu$ and $\theta$ then scales linearly with $\beta$.

The validation result also suggests that $\epsilon$ can be modeled as a Gaussian random variable. Thus we let $A=\mu-3\theta$ and $B=\mu+3\theta$ so that it contains more than 99\% of the probability mass and results in reasonably good numerical accuracy and approximation \cite{PD99}. The spot transit capacity $C$ is set to $(0.4+\beta)$ times the aggregated regular transit demand $\bar{d}(\bar{p})$ throughout the evaluation. For a backbone with peak utilization of 50\%, the underutilized capacity equals $\bar{d}(\bar{p})$. 50\%--110\% of this underutilized capacity is thus safe to be used for spot transit. We believe such setting represents a typical operating environment of spot transit. One can readily verify that $B<C$ holds as assumed in Sec.~\ref{subsec:profit_model}.


    

Note that we have used the simplest prediction algorithm, and the result is therefore only a lower bound. With more complex algorithms one can obtain more accurate prediction for a longer time window \cite{PTZD03}, which is beyond the focus of this paper.    
The prediction techniques and the line of reasoning apply to both elastic and inelastic traffic. We emphasize that our demand uncertainty model is dependent only on the observation that demand at a tier-1 ISP is highly predictable.

\begin{figure*}[tbp]
    \begin{minipage}[t]{0.33\linewidth}
    \centering
    \includegraphics[width=1\linewidth]{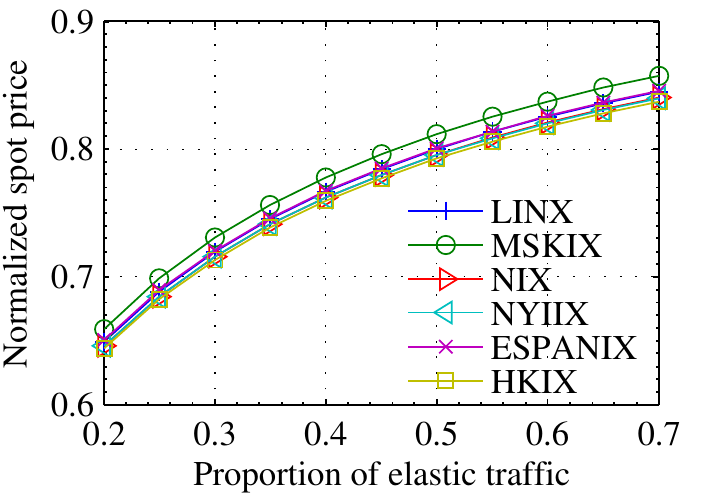}
    \caption{Optimal spot price with iso-elastic demand. }
    \label{fig:ced-price}
    \end{minipage}
    \begin{minipage}[t]{0.33\linewidth}
    \centering
    \includegraphics[width=1\linewidth]{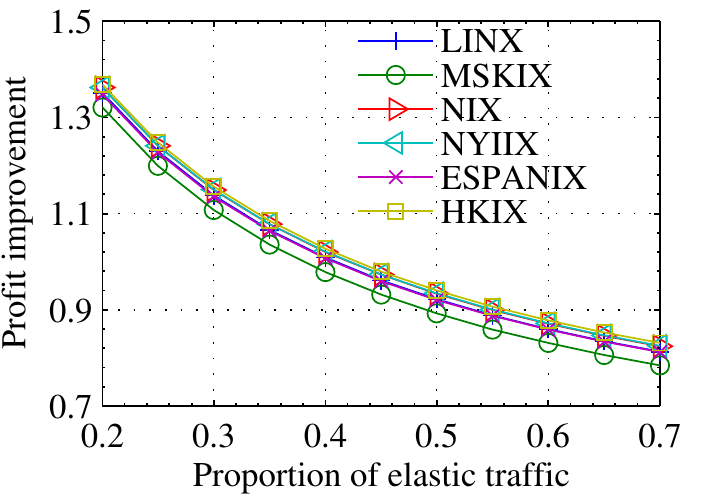}
    \caption{Profit improvement with iso-elastic demand.}
    \label{fig:ced-profitimprov}
    \end{minipage}
    \begin{minipage}[t]{0.33\linewidth}
    \centering
    \includegraphics[width=1\linewidth]{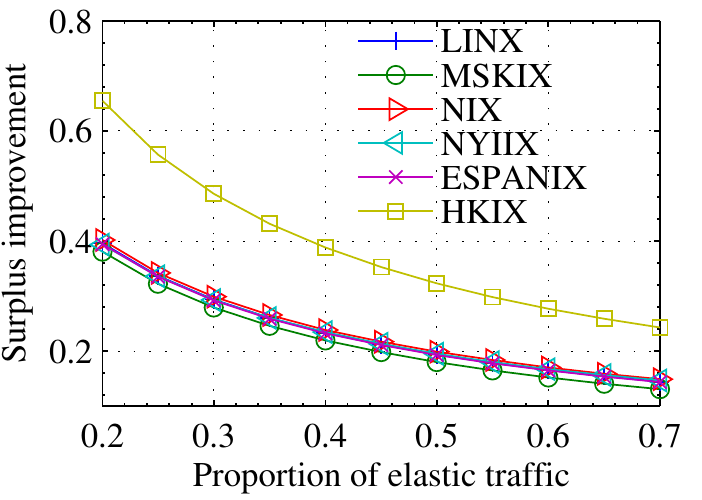}
    \caption{Surplus improvement with iso-elastic demand.}
    \label{fig:ced-surplusimprov}
    \end{minipage} 
\end{figure*}
\begin{figure*}[tbp]
    \begin{minipage}[t]{0.33\linewidth}
    \centering
    \includegraphics[width=1\linewidth]{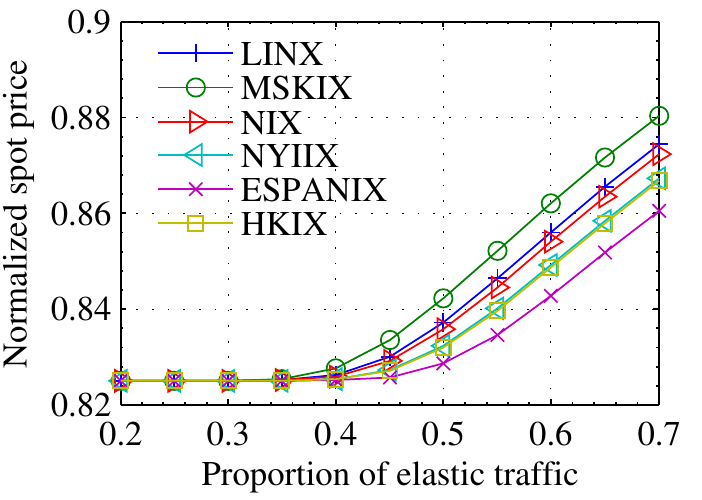}
    \caption{Optimal spot price with linear demand. }
    \label{fig:linear-price}
    \end{minipage}
    \begin{minipage}[t]{0.33\linewidth}
    \centering
    \includegraphics[width=1\linewidth]{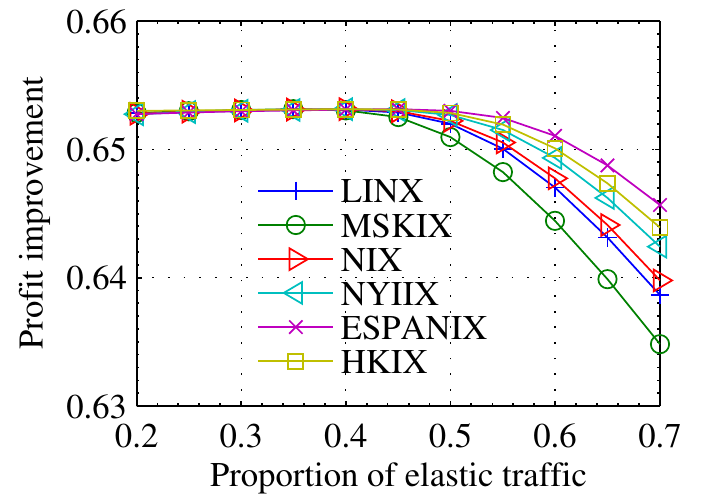}
    \caption{Profit improvement with linear demand.}
    \label{fig:linear-profitimprov}
    \end{minipage}
    \begin{minipage}[t]{0.33\linewidth}
    \centering
    \includegraphics[width=1\linewidth]{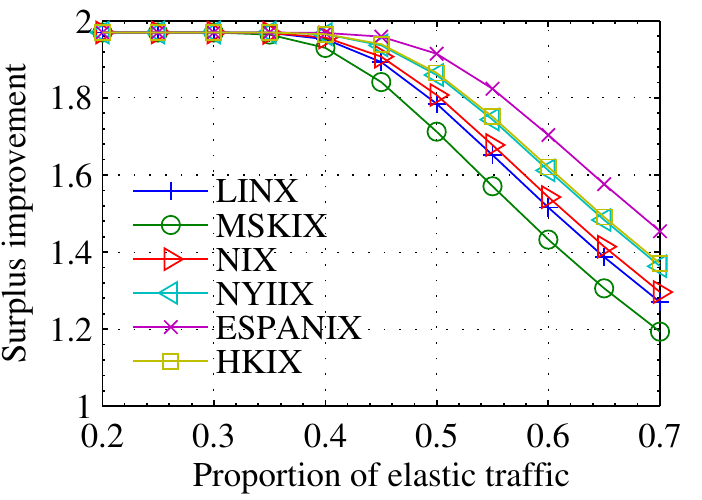}
    \caption{Surplus improvement with linear demand.}
    \label{fig:linear-surplusimprov}
    \end{minipage} 
\end{figure*}

\subsection{Obtaining cost and demand parameters from data}
\label{sec:demandparameters}

After validating the model, the next key step is to obtain cost and demand parameters in our model. 
Our analysis in Sec.~\ref{sec:analysis} is applicable to general demand functions. In the evaluation we use two common demand functions, iso-elastic and linear demand as in Sec.~\ref{subsubsec:d_p}. To derive model parameters for them, we use the Q2 2011 median GigE transit price in New York, London, and Hong Kong published in \cite{T11} as the regular transit price $\bar{p}$ in America, Europe, and Asia. Table~\ref{table:prices} lists the price data. 

First, to obtain the cost of regular transit $\bar{r}$, since now we know the regular transit demand $\bar{d}(\bar{p})$ and $\bar{p}$, assuming ISPs are rational and profit-maximizing, from \eqref{eqn:pbar} we can obtain 
\begin{equation}\label{eqn:barr_ced}
  \bar{r} = \bar{p}\left(1-\frac{1}{\bar{\alpha}}\right)
\end{equation}
for iso-elastic demand with \eqref{eqn:ced}, and
\begin{equation}\label{eqn:barr_linear}
  \bar{r} = \bar{p}- \frac{\bar{d}(\bar{p})}{\bar{\alpha}}
\end{equation}
for linear demand with \eqref{eqn:lineard}. Studies have shown that the more elastic residential Internet traffic has an elasticity of around 2.7 for cable and DSL \cite{CSYZ07}. 
We thus assume the demand elasticity $\bar{\alpha}=2$ in \eqref{eqn:barr_ced} for aggregated traffic, i.e.~d$\bar{r}=0.5\bar{p}$. Since $\bar{r}$ is invariant across two demand models, we can substitute $\bar{r}$ into \eqref{eqn:barr_linear} to obtain $\bar{\alpha}$ for linear demand in different continents.

To obtain demand parameters, we first multiply $\beta\in [0.2,0.7]$ to $\bar{d}(\bar{p})$ to calculate the elastic traffic demand at the regular price $d(\bar{p})$ from the IXP dataset. Then from \eqref{eqn:ced} and \eqref{eqn:lineard} we know that as long as the demand sensitivity parameter $\alpha$ is known we can obtain the base demand  
\begin{equation}
  v = d(\bar{p})\cdot\bar{p}^{\alpha}
\end{equation}
for iso-elastic demand and 
\begin{equation}
  v = d(\bar{p})+\alpha\bar{p}
\end{equation}
for linear demand. The entire elastic demand curve $d(p)$ can then be obtained. We use a range of values for $\gamma>1$ to control the relative elasticity of spot transit versus regular transit, in order to evaluate the effect of elasticity on the benefits of spot transit. For iso-elastic demand, $\alpha$ corresponds directly to elasticity, and $\alpha = \gamma \bar{\alpha}$ so that spot transit is $\gamma$ more sensitive. For linear demand, since $\alpha$ also affects the demand magnitude, we have to first scale it down by $\beta$, i.e.~$\alpha = \beta\cdot\gamma\bar{\alpha}$.  

\begin{table}
      \centering
    \caption{The regular transit prices in major Internet exchange locations \cite{T11}.}
    \begin{tabular}{| c | c |}
    \hline 
    Location & Price $\bar{p}$ (\$USD/Mbps)\\ \hline
        London  & 7.5\\
        New York & 7\\ 
        Hong Kong & 22\\
    \hline  
    \end{tabular}
\label{table:prices}
\end{table}

We stress that the purpose of evaluation is to verify the analysis in Sec.~\ref{sec:analysis} and gain insights on the potential of spot transit in a realistic setting. We do not claim the numerical accuracy of the results obtained here for tier-1 ISPs. The exact pricing and monetary benefits heavily depend on various factors and can only be calculated on a case-by-case basis.

\subsection{Overall benefits}
\label{sec:overall_benefits}

\begin{figure}[htbp]
   \begin{minipage}[t]{0.49\linewidth}
    \centering
    \includegraphics[width=1\linewidth]{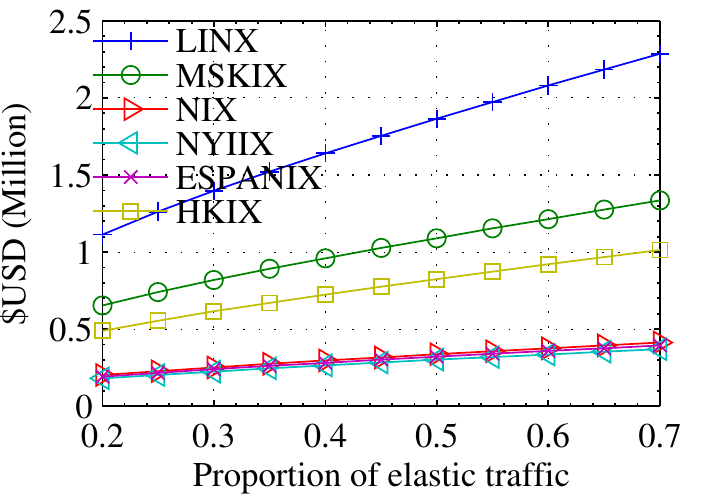}
    \caption{Profit gain of spot transit with iso-elastic demand.}
    \label{fig:ced-profit}
    \end{minipage}
    \begin{minipage}[t]{0.49\linewidth}
    \centering
    \includegraphics[width=1\linewidth]{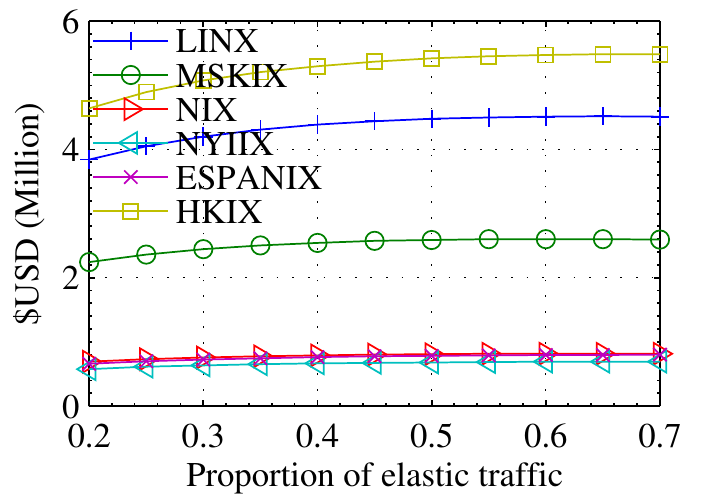}
    \caption{Surplus gain of spot transit with iso-elastic demand.}
    \label{fig:ced-surplus}
    \end{minipage} 
\end{figure}

\begin{figure}[htbp]
    \begin{minipage}[t]{0.49\linewidth}
    \centering
    \includegraphics[width=1\linewidth]{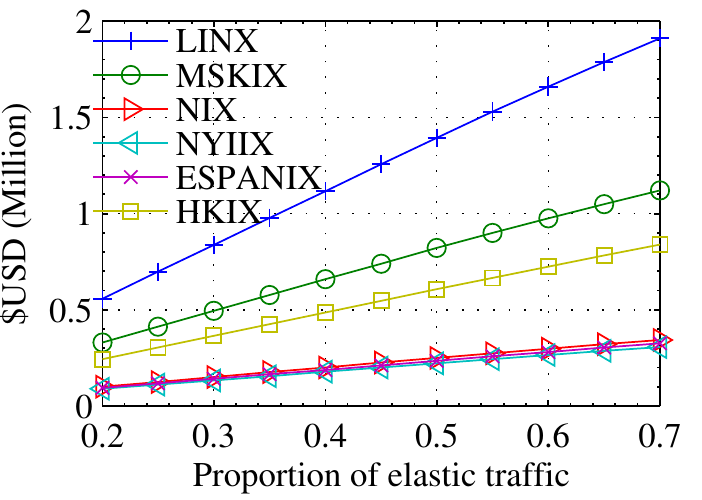}
    \caption{Profit gain of spot transit with linear demand.}
    \label{fig:linear-profit}
    \end{minipage}
    \begin{minipage}[t]{0.49\linewidth}
    \centering
    \includegraphics[width=1\linewidth]{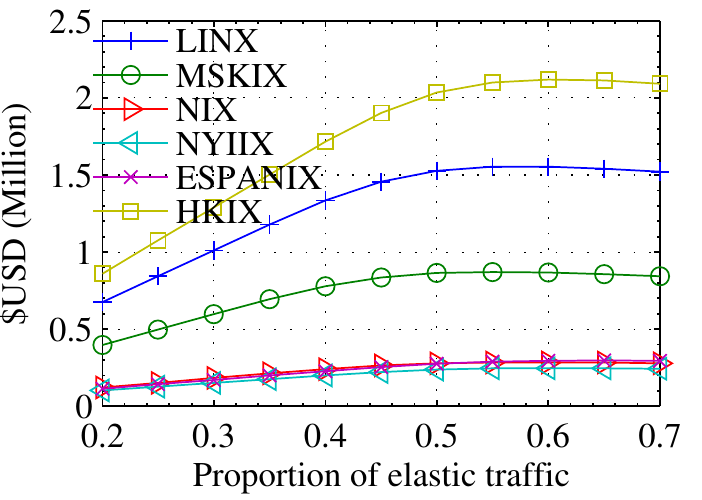}
    \caption{Surplus gain of spot transit with linear demand.}
    \label{fig:linear-surplus}
    \end{minipage} 
\end{figure}

First and foremost, we evaluate the overall benefits of spot transit with typical parameter setting. For both demand models, we set $\gamma=1.25$, $r=0.5\bar{r}$, and $m=\bar{p}$ so that elastic traffic, i.e. spot transit demand, is 1.25 times more sensitive than regular transit, cost is half of the regular cost, and overflow penalty is equal to regular price. $\beta=[0.2, 0.7]$, and the capacity $C=(0.4+\beta)\bar{d}(\bar{p})$. For a backbone with peak utilization of 50\%, the underutilized capacity equals $\bar{d}(\bar{p})$. 50\%-110\% of this underutilized capacity is thus safe to be used for spot transit.
We believe such setting represents a typical operating environment of spot transit. Figure~\ref{fig:ced-price}-\ref{fig:linear-surplus} show the evaluation results of all 6 IXPs for both demand models. 

Figure~\ref{fig:ced-price} and \ref{fig:linear-price} plot the normalized spot transit price $\frac{p*}{\bar{p}}$ for both demand models. Observe that spot transit is offered at a discount, ranging from more than 30\% to 15\%. This demonstrates the price advantage of spot transit. Price increases with $\beta$, suggesting that as relative proportion of elastic traffic increases, demand overflow probability increases since cost, penalty, and elasticity does not change with $\beta$ here. Figure~\ref{fig:ced-profitimprov} and \ref{fig:linear-profitimprov} show the profit improvement of spot transit. We can see that spot transit significantly improves tier-1 ISP's profit, by 80\%-130\% with iso-elastic demand, and by 63\%-65\% with linear demand. Same observation can be made in terms of consumer surplus as shown in Figure~\ref{fig:ced-surplus} and \ref{fig:linear-surplus}. Spot transit improves surplus by 10\%-40\% with iso-elastic demand, and by 120\%-200\% with linear demand. Since price increases with $\beta$, the improvement decreases as a result.


One may wonder at this point, what is the exact dollar amount of spot transit's benefits? Figure~\ref{fig:ced-profit} and \ref{fig:linear-surplus} plot the absolute profit and surplus gain for iso-elastic demand, and Figure~\ref{fig:ced-surplus} and \ref{fig:linear-profit} plot the absolute gains for linear demand, respectively. The profit gain stands more than \$1 million for large IXPs like LINX, and in the order of hundreds of thousands of dollars for smaller ones like NIX, NYIIX and ESPANIX for both models. The absolute surplus gain depends more on the shape of demand curve, and is more salient with iso-elastic demand that allows price to go to infinity. Again the gain is in the order of million dollars for large IXPs, and hundreds of thousands of dollars for smaller ones. Thus, our evaluation not only confirms the qualitative analysis in Sec.~\ref{sec:efficiency}, but also quantitatively shows that spot transit offers significant financial incentives with more than \$1 million dollars of gains possible (monthly) for both tier-1 ISPs and transit customers, depending on the size of the ISP. 

Another interesting observation from the results is that, the smallest IXP, HKIX, enjoys more dramatic performance improvement than others despite its relatively small scale, especially in terms of consumer surplus. The reason is that the regular transit price in Asia and other more remote areas of Internet is much higher than other regions, resulting in a much larger profit margin in \$/Mbps for tier-1 ISPs. Thus, with the same price discount, its profit and surplus gains are more profound than bigger IXPs in major cities of Europe and America. This illustrates that spot transit is potentially more attractive in ``remote'' regions of Internet where regular transit is much more expensive.

\subsection{Pricing analysis}
\label{sec:pricing_evaluation}

\begin{figure*}[htbp]
    \begin{minipage}[t]{0.33\linewidth}
    \centering
    \includegraphics[width=1\linewidth]{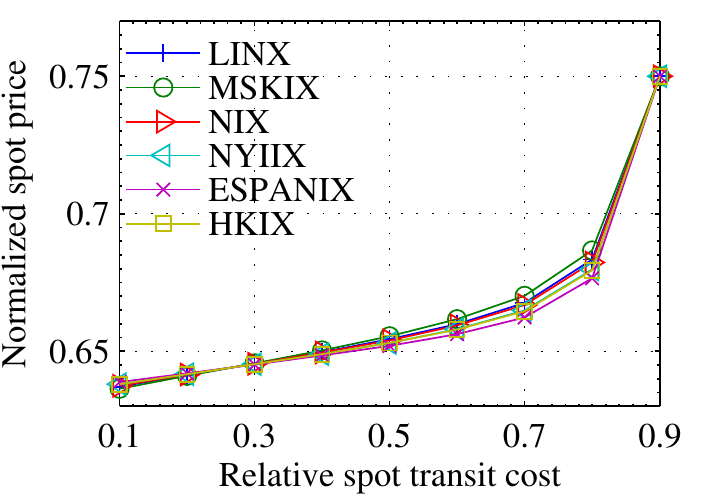}
    \caption{The effect of $r$ on spot price with iso-elastic demand. }
    \label{fig:r-ced-price}
    \end{minipage}
    \hspace{-0cm}
    \begin{minipage}[t]{0.33\linewidth}
    \centering
    \includegraphics[width=1\linewidth]{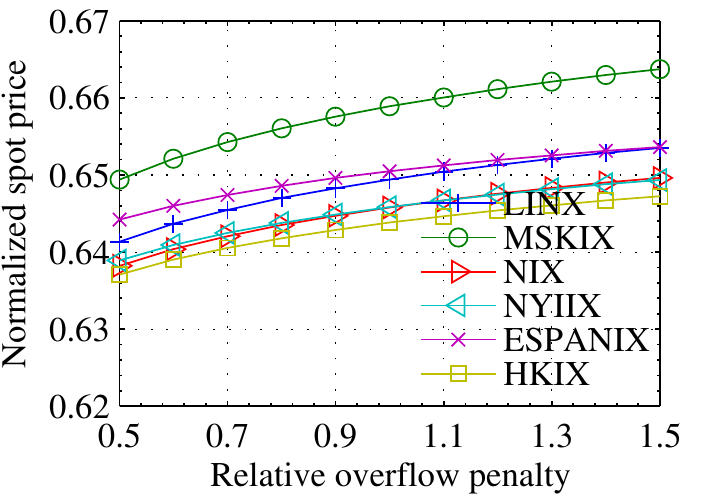}
    \caption{The effect of $m$ on spot price with iso-elastic demand.}
    \label{fig:m-ced-price}
    \end{minipage}
    \begin{minipage}[t]{0.33\linewidth}
    \centering
    \includegraphics[width=1\linewidth]{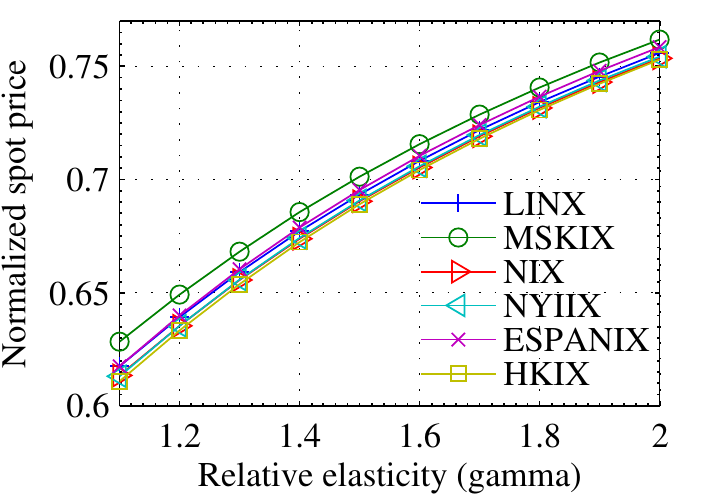}
    \caption{The effect of $\gamma$ on spot price with iso-elastic demand.}
    \label{fig:gamma-ced-price}
    \end{minipage}
\end{figure*}

Now we the effect of various parameters on spot transit pricing. We choose to present results with iso-elastic demand since both models lead to similar conclusions.

Figure~\ref{fig:r-ced-price}-\ref{fig:gamma-ced-price} show how spot transit price is affected by cost $r$, penalty $m$, and elasticity $\alpha$, respectively. We vary the relative cost $r/\bar{r}$ between 0.1 and 0.9 with $m=\bar{p}$ and $\gamma=1.25$, relative penalty $m/\bar{p}$ between 0.5 and 1.5 with $r=0.5\bar{r}$ and $\gamma=1.25$, and relative elasticity $\gamma$ between 1.1 and 2 with $r=0.5\bar{r}$ and $m=\bar{p}$. Other parameter settings remain the same as in the previous section. We observe that spot price increases with all of the three factors, as expected from Theorem~\ref{thm:monopoly}. Price is less sensitive to penalty $m$ compared to cost and elasticity, since penalty is only imposed on the overflown portion of demand. Also from Figure~\ref{fig:r-ced-price} we can see that when $r\ge0.4\bar{r}$, $\bar{r}-r\le 0.6\bar{r}=0.3\bar{p} < 0.5m(1-\frac{1}{2*1.25})=0.3\bar{p}$, spot transit price is still offered at more than 15\% discount. This confirms our discussion of Theorem~\ref{thm:p*_properties} in Sec.~\ref{sec:pricing} that even when the condition $r\le \bar{r}-0.5m(1-\sigma(p^*)^{-1})$ is not satisfied, i.e. when the cost difference is not large, spot transit can still be much cheaper than regular transit, and improves the overall efficiency of the market as proved in Theorem~\ref{thm:welfare}.

\subsection{Sensitivity analysis}
\label{sec:sensitivity}

We have studied how cost $r$, penalty $m$, and elasticity $\alpha$ affect spot transit pricing. In this section, we analyze how these model parameters affect the profit and surplus improvement of spot transit. 

\begin{figure}[htbp]
    \begin{minipage}[t]{0.49\linewidth}
    \centering
    \includegraphics[width=1\linewidth]{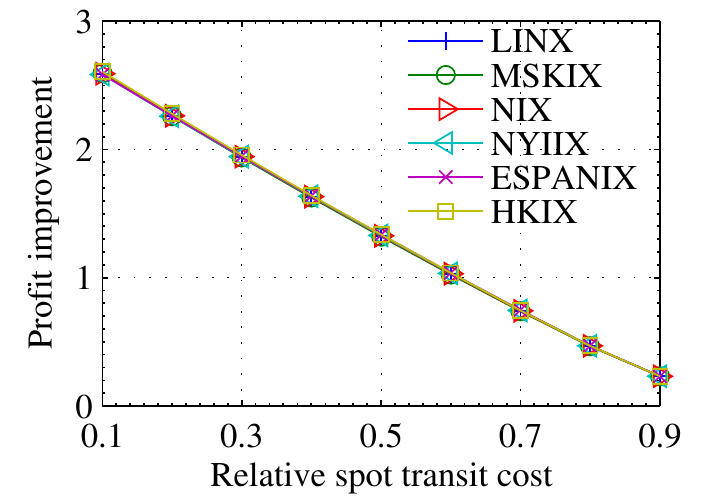}
    \caption{Profit improvement vs. $r$ with iso-elastic demand. }
    \label{fig:r-ced-profitimprov}
    \end{minipage}
    \hspace{-0cm}
    \begin{minipage}[t]{0.49\linewidth}
    \centering
    \includegraphics[width=1\linewidth]{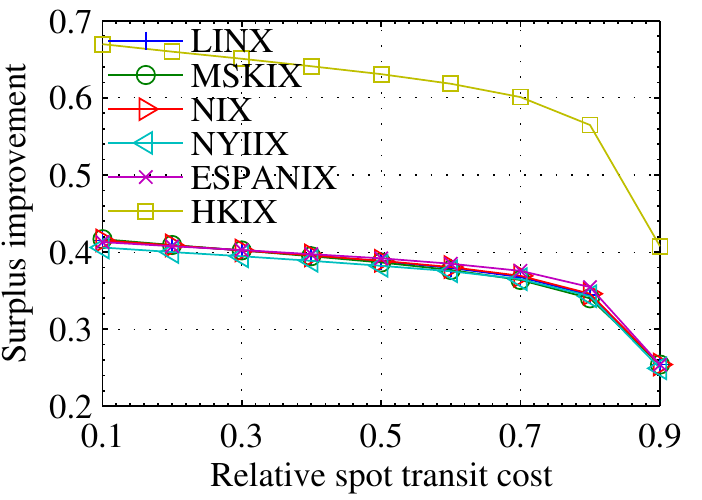}
    \caption{Surplus improvement vs. $r$ with iso-elastic demand.}
    \label{fig:r-ced-surplusimprov}
    \end{minipage}
\end{figure}

\begin{figure}[htbp]
    \begin{minipage}[t]{0.49\linewidth}
    \centering
    \includegraphics[width=1\linewidth]{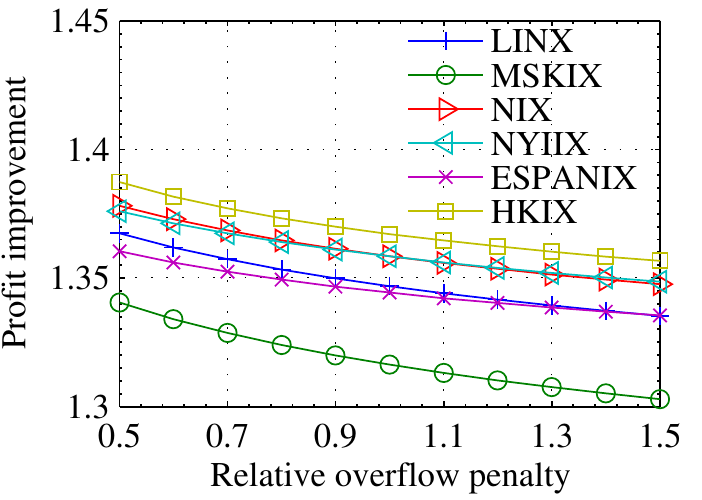}
    \caption{Profit improvement vs. $m$ with iso-elastic demand. }
    \label{fig:m-ced-profitimprov}
    \end{minipage}
    \hspace{-0cm}
    \begin{minipage}[t]{0.49\linewidth}
    \centering
    \includegraphics[width=1\linewidth]{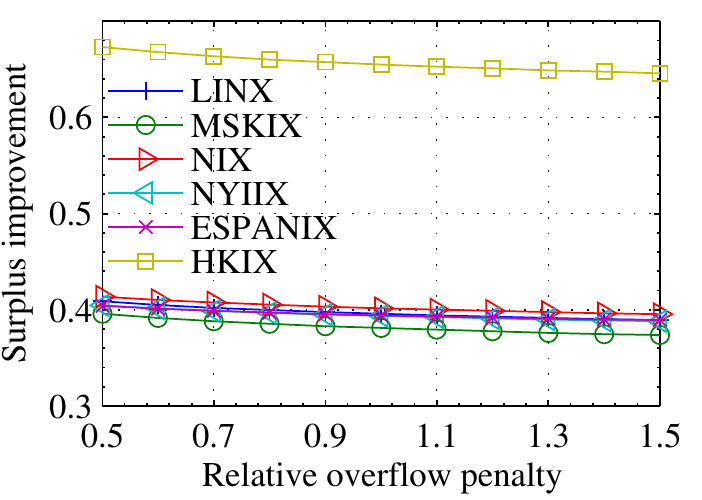}
    \caption{Surplus improvement vs. $m$ with iso-elastic demand.}
    \label{fig:m-ced-surplusimprov}
    \end{minipage}
\end{figure}

First we vary $r$ and $m$ individually as we did in the previous section, and plot the profit and surplus improvement with iso-elastic demand in Figure~\ref{fig:r-ced-profitimprov}-\ref{fig:m-ced-surplusimprov}. Observe that as cost $r$ and penalty $m$ increase, both profit and surplus drop which is intuitive to understand. Spot transit is able to provide positive improvement even when $r=0.9\bar{r}$ or $m=1.5\bar{p}$. Results of linear demand are similar and not presented.

\vspace{-4mm}

\begin{figure}[htbp]
    \begin{minipage}[t]{0.49\linewidth}
    \centering
    \includegraphics[width=1\linewidth]{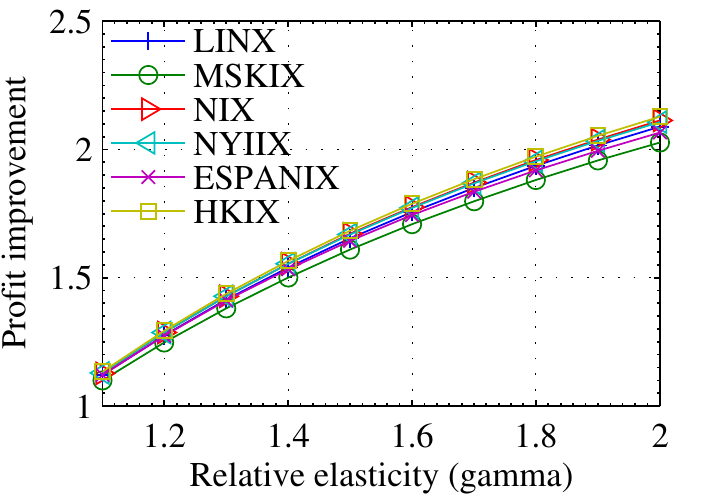}
    \caption{Profit improvement vs. $\gamma$ with iso-elastic demand. }
    \label{fig:gamma-ced-profitimprov}
    \end{minipage}
    \hspace{-0cm}
    \begin{minipage}[t]{0.49\linewidth}
    \centering
    \includegraphics[width=1\linewidth]{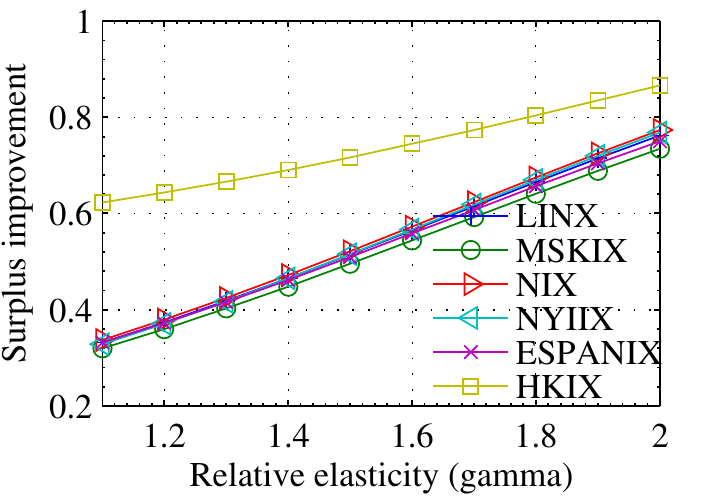}
    \caption{Surplus improvement vs. $\gamma$ with iso-elastic demand.}
    \label{fig:gamma-ced-surplusimprov}
    \end{minipage}
\end{figure}

We then vary elasticity $\alpha$ by varying $\gamma$ between 1.1 and 2. Figure~\ref{fig:gamma-ced-profitimprov}-\ref{fig:gamma-ced-surplusimprov} show the corresponding profit and surplus improvement. We can see that although a larger $\alpha$ increases spot transit prices, it also increases the profit and surplus gains at the same time, which is in sharp contrary to the results of cost and penalty. The reason for the discrepancy is that increasing $\alpha$ does not change cost and penalty of spot transit, and as a result the profit margin is actually increased due to the price increase. From the numerical result we also observe that demand at $p^*$ also increases despite the price increase because the elasticity now is larger. The overall effect of increased elasticity therefore is positive, in spite of slightly increased revenue loss due to demand overflow either.   

\begin{figure}[htbp]
    \begin{minipage}[t]{0.49\linewidth}
    \centering
    \includegraphics[width=1\linewidth]{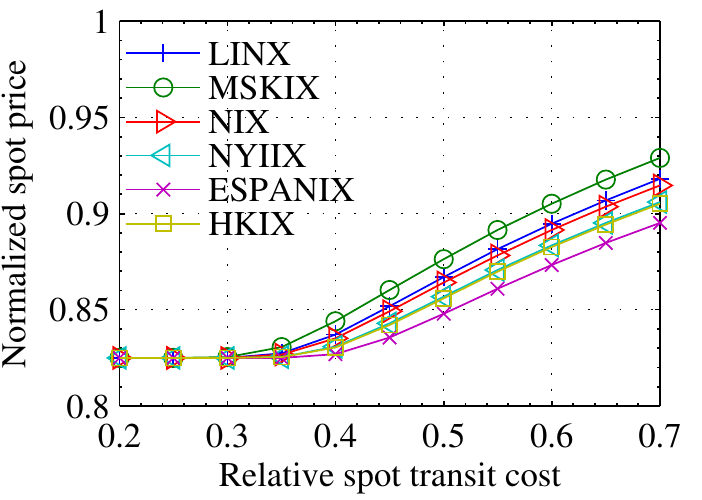}
\vspace{-4mm}
    \caption{Worst-case price with iso-elastic demand. }
    \label{fig:worst-ced-price}
    \end{minipage}
    \begin{minipage}[t]{0.49\linewidth}
    \centering
    \includegraphics[width=1\linewidth]{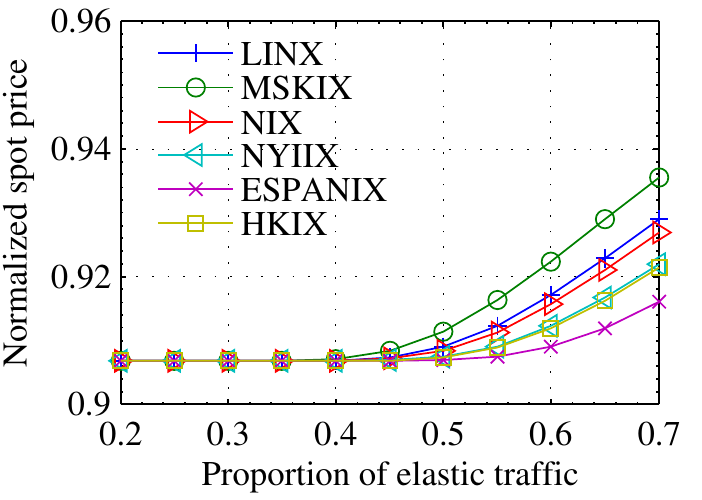}
\vspace{-4mm}
    \caption{Worst-case price with linear demand.}
    \label{fig:worst-linear-price}
    \end{minipage}
\end{figure}

\begin{figure}[!hptb]
    \begin{minipage}[t]{0.49\linewidth}
    \centering
    \includegraphics[width=1\linewidth]{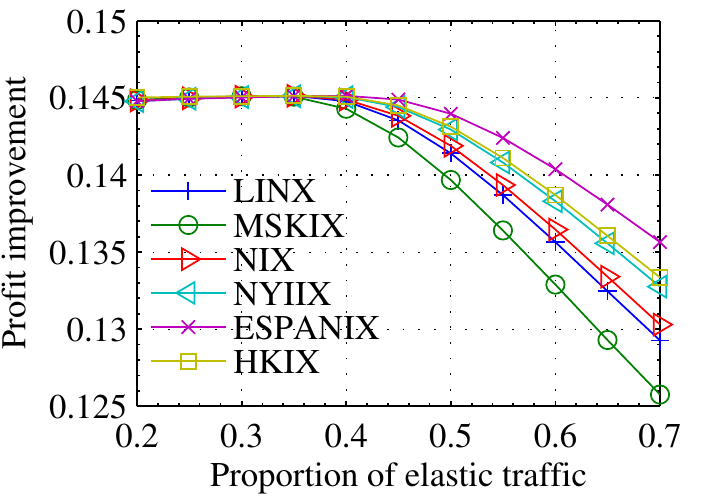}
    \caption{Worst-case profit improvement with iso-elastic demand. }
    \label{fig:worst-ced-profitimprov}
    \end{minipage}
    \begin{minipage}[t]{0.49\linewidth}
    \centering
    \includegraphics[width=1\linewidth]{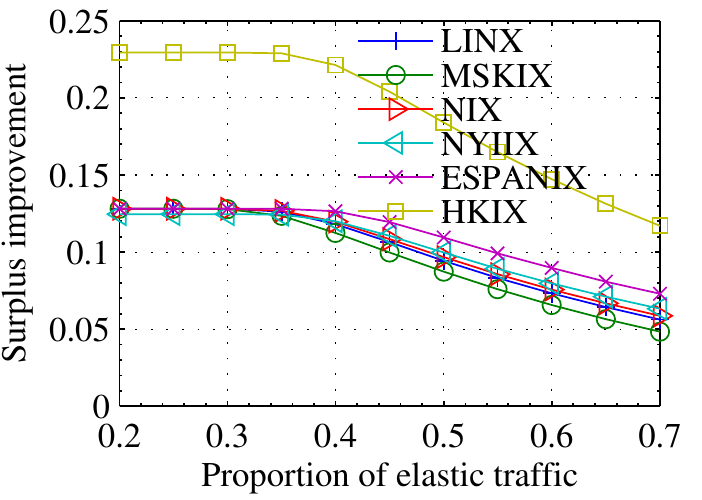}
    \caption{Worst-case surplus improvement with iso-elastic demand.}
    \label{fig:worst-ced-surplusimprov}
    \end{minipage}
\end{figure}

\begin{figure}[hpbt]
    \begin{minipage}[t]{0.49\linewidth}
    \centering
    \includegraphics[width=1\linewidth]{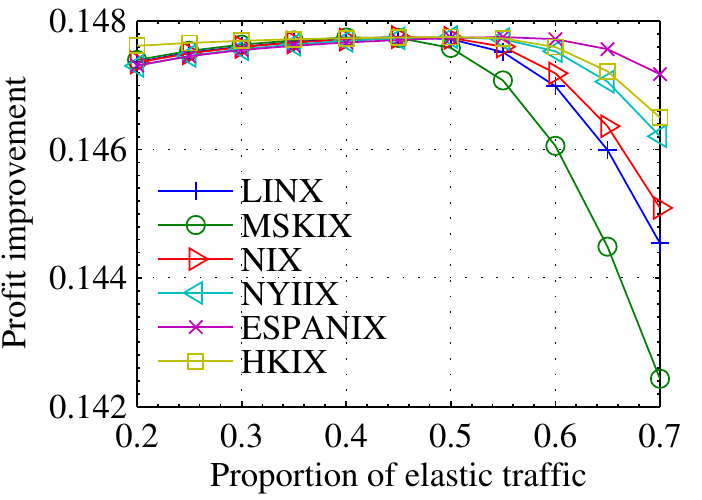}
    \caption{Worst-case profit improvement with linear demand. }
    \label{fig:worst-linear-profitimprov}
    \end{minipage}
    \begin{minipage}[t]{0.49\linewidth}
    \centering
    \includegraphics[width=1\linewidth]{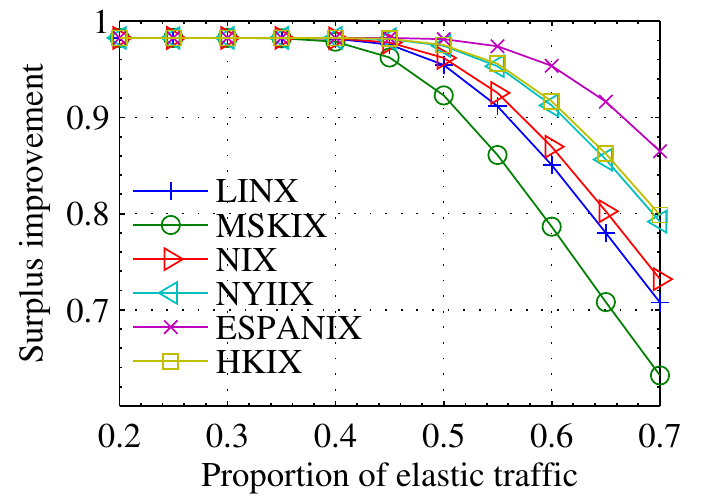}
    \caption{Worst-case surplus improvement with linear demand.}
    \label{fig:worst-linear-surplusimprov}
    \end{minipage}
\end{figure}

Finally, we study a {\em worst-case} scenario, where all the three parameters are deliberately chosen to represent the worst operating environment with high cost, high penalty, and low elasticity for spot transit. Specifically, we let $r=0.9\bar{r}$, $m=1.5\bar{p}$, and $\gamma=1.1$, and plot the spot transit price, profit and surplus improvement with varying $\beta$ in Figure~\ref{fig:worst-ced-price}-\ref{fig:worst-linear-surplusimprov}. In other words, these figures represent the {\em minimum} improvement over a range of parameter values. 

We can see that with the worst combination of parameters, spot transit is still slightly cheaper than regular transit for both demand models. The profit improvement stands above 10\%, and the surplus improvement is 5\% with iso-elastic demand and more than 60\% with linear demand. The results clear demonstrates that the advantage of spot transit is robust against a wide range of parameter values, and spot transit can be expected to provide significant gains in a typical operating environment.


%% file: discussion.tex
\section{Discussion of Feasibility}
\label{sec:discussion}

We have compellingly demonstrated, through both theoretical analysis and empirical evaluation based on real-world traffic and price data, that spot transit provides significant financial benefits to tier-1 ISPs and transit customers. However, profitability alone does not guarantee the feasibility of spot transit. In this section, we examine some practical aspects that we believe are important to the establishment of such a new transit settlement market. 

\subsection{Market infrastructure}
\label{sec:infrastructure}
In the spot transit market, customers can purchase and utilize spot transit on-demand from tier-1 ISPs. This requires a physically connected infrastructure amongst all customers and the ISP. 
We believe this seemingly daunting task has, to a large extent, been solved by the proliferation of network exchange facilities, such as Internet Exchange Points (IXP) and Network Access Points (NAP), across the world. They host hundreds of ISPs each already, including tier-1 ISPs \cite{IXP}. Though IXPs carry mostly peering traffic for now, the infrastructure can certainly be utilized to support spot transit with little additional cost. Such a public infrastructure enables tier-1 ISPs to support any customers present in the IXP, and customers to flexibly switch between spot transit providers. By supporting spot transit IXPs also diversify and expand its business line.

In case the tier-1 ISP or the transit customer is not present in an IXP, it is highly likely that a private link exists between the two for carrying regular transit. Spot transit can then be provided over the existing private link. Even in the extremely rare case that a new link has to be set up, such a one-time cost is expected to be rather insignificant compared to the long-term benefits of spot transit.

\subsection{Inter-domain routing and billing}
\label{sec:routing}
The introduction of spot transit traffic does not pose technical challenges or complications for inter-domain routing of BGP (Border Gateway Protocol). Though it shares the same network backbone with regular transit traffic, spot transit traffic can be easily identified and managed by a designated AS (Autonomous System) number acquired by the tier-1 ISP. The ISP advertises routes with the designated AS number when it supports spot transit, and stops doing so when it does not wish to carry spot transit due to say insufficient capacity or other considerations. Billing is also straightforward by tracking traffic destined to the designated AS number. No contract is required and only the used amount of bandwidth is billed based on 95-th percentile billing or any other suitable method. The spot transit bill can easily be combined with the regular transit bill if the customer uses both from the same tier-1 ISP.

\subsection{Market cannibalization}
\label{sec:strategics}
One may be concerned that the tier-1 ISP's regular transit business would be negatively impacted by offering spot transit customers, resulting in the so-called market cannibalization \cite{market-cann}. We have already shown that, the spot transit profit is significantly larger than the profit collected from serving the elastic traffic with regular transit due to an increase of demand as a result of price reduction. The demand increase does not necessarily translate to a decrease of regular transit demand. In fact we do not expect spot transit to be suitable for the relatively inelastic traffic which finds regular transit with SLAs more reliable. 

The demand increase with spot transit can be explained by at least two factors: competition with peering and transit reselling. The low price and on-demand feature of spot transit can make it more appealing than peering or paid peering, considering the performance benefits and network reachability provided by the tier-1 backbone. An ISP that relies mostly on (paid) peering for cost reasons can utilize spot transit for a portion of its elastic traffic with much better reachability. Further, small ISPs usually find it difficult to purchase transit directly from tier-1 ISPs due to the minimum committed data rate requirement. They purchase transit from medium size ISPs that buy transit at bulk and resell to these small customers. With spot transit, tier-1 ISPs are able to collect additional profits from small ISPs by bypassing the transit resellers in the middle. In all, we expect that spot transit {\em compliments} rather than cannibalizes the traditional transit business of a tier-1 ISP.

%% file: related.tex
\section{Related Work}
\label{sec:related}


An extensive literature exists on the Internet transit market in both networking and economics. Two aspects are particularly related to our work: optimal pricing design, and novel market approaches for Internet transit. 

Internet broadband access pricing generally is designed and computed to optimize revenue, social welfare, or performance. \cite{O00} argues that the predominant flat-rate pricing structure for selling retail Internet access encourages waste and is incompatible with service differentiation. \cite{KDV08,SSOA08} study the benefits of usage-based pricing and argue that, with price differentiation, one can use resources more efficiently. \cite{O99,CWC10} study Paris Metro Pricing in which service differentiation and congestion control are autonomously achieved by charging different prices for different service tiers that share the same infrastructure. Time is another dimension to unbundle connectivity. Hande {\em et al.}~\cite{HCCZ10} characterize the economic loss due to the ISP's inability or unwillingness to price broadband access based on time of day. Jiang {\em et al.}~\cite{JHC11} study the optimal time-dependent prices for an ISP selling broadband access based on solving optimization offline with traffic estimates. 

Our work is different in that we study pricing of spot transit, a new market for Internet transit. \cite{MCLM08} proposes a Shapley value based cooperative settlement between content, transit, and eyeball ISPs. The focus is on performance and optimality of the Internet ecosystem with selfish ISPs through fair and efficient profit sharing. \cite{VFJV08} proposes a clean-slate market structure and routing protocol for exchange of Internet paths. \cite{VLFJ11,SCG11,CG12} are more related to our work. \cite{VLFJ11} studies tiered pricing based on packet destinations and routing costs for selling Internet transit. It is shown that a few tiers is enough to capture the optimal profit gain for a tier-1 ISP. From the customer's perspective, \cite{SCG11} proposes to use Tuangou (group buying), and \cite{CG12} innovates T4P (transit for peering) that provides partial transit to peering partners, to reduce transit costs.

Spot transit is orthogonal to these novel settlement schemes, and compliments them in that tiered pricing, Tuangou, and T4P can be readily applied in spot transit market just like they are used in the regular transit market. The benefits of spot transit, as mentioned, depend on the very characteristics of elastic traffic instead of the specifics of settlement details. We use a simple model derived from \cite{M59,PD99} where demand randomness is modeled additively. The technical distinction is clear: we use a general demand function and analyze the profit and surplus improvement, while \cite{M59,PD99} only study pricing based on a linear demand function.
 


%% file: conclusion.tex
\section{Concluding Remarks}
\label{sec:conclusion}

In this paper, we advocate to create a \emph{spot} transit market, where under-utilized backbone capacity is offered at discounted to serve elastic traffic, and transit customers can purchase transit on-demand. We systematically studied the pricing and economical benefits of spot transit. Through both theoretical analysis and empirical evaluation with real-world price and traffic data, we demonstrated that significant profit and surplus improvement can be generally expected from the spot transit market. The gains are also robust for a wide range of parameter settings. Given the potential economical benefits, we believe spot transit will encourage many entities to engage in this new market.

We conclude the paper by pointing out some interesting open problems with the introduction of spot transit. For example, how can the tier-1 ISP use smart traffic engineering algorithms to provide better performance isolation between the regular and spot transit traffic? How do we quantify the effect of spot transit on novel settlement schemes such as paid peering? How would it change the Internet AS level topology, and the entire ecosystem?


%% file: appendix.tex
\appendix

\section{Proof of Lemma~1}
\label{app:lem:quasi}

The risk-free profit $\Phi(p)=(p-r)d(p)$ is quasiconcave, since $p-r$ is monotonic and thus quasiconcave, $d(p)$ is concave and thus quasiconcave, and the product of two quasiconcave functions are quasiconcave. From \eqref{eqn:profit} and \eqref{eqn:profit_short}, the first-order derivative of the profit loss function $\Lambda(p)$ is $d'(p)m\int^B_{C-d(p)} f(u)\ud u$, which is increasing in $p$. Thus $-\Lambda(p)$ is concave in $p$. Since $E[R(p)]=\Phi(p)-\Lambda(p)$, it can be readily shown that the sum of a quasiconcave function and a concave function is quasiconcave.

\section{Proof of Theorem~2}
\label{app:thm:p*}

Since $\bar{d}(p)$ includes both inelastic and elastic traffic demand, its elasticity is smaller, i.e. $\bar{\sigma}(p)<\sigma(p)$ for any $p$. The first-order condition of \eqref{eqn:total_profit} amounts to 
\begin{equation}\label{eqn:pbar}
    \bar{p} = \bar{r} - \frac{\bar{d}(\bar{p})}{\bar{d}'(\bar{p})} \Rightarrow \bar{p} = \frac{\bar{r}}{1-\bar{\sigma}(\bar{p})^{-1}}
\end{equation}
by substituting \eqref{eqn:elasticity}. This implies that $1<\bar{\sigma}(\bar{p})$. At the optimal spot price $p^*$, $d(p^*)<C$ always holds as discussed in Sec.~\ref{subsec:profit_model}. Thus substituting \eqref{eqn:elasticity} into \eqref{eqn:p*}, and applying the one-sided Chebyshev Inequality (Chebyshev-Cantelli Inequality) to upper bound $\Pr(\epsilon>C-d(p^*))$,
\begin{gather*}
	p^* < \frac{r}{1-\sigma(p^*)^{-1}} + ma,\ \text{where }a=\frac{\theta^2}{\theta^2+(C-d(p^*)-\mu)^2}.  
\end{gather*}
$\mu$ and $\theta$ are the mean and standard deviation of $\epsilon$, respectively.
Now assume that $p^*\ge\bar{p}$, which implies
\begin{equation*}
    \frac{\bar{r}}{1-\bar{\sigma}(\bar{p})^{-1}} < \frac{r+ma\left(1-\sigma(p^*)^{-1}\right)}{1-\sigma(p^*)^{-1}}.
\end{equation*}
$1<\bar{\sigma}(\bar{p}) \le \bar{\sigma}(p^*) < \sigma(p^*)$ by \eqref{eqn:sigma'}, and $0<1-\bar{\sigma}(\bar{p})^{-1} < 1- \sigma(p^*)^{-1}$. Thus,
\begin{equation*}
    \bar{r} < r + ma\left(1-\sigma(p^*)^{-1}\right),
\end{equation*}
which contradicts with condition \eqref{eqn:price_condition}.

\section{Proof of Lemma~3}
\label{app:lem:profit}
Substituting \eqref{eqn:p*} into \eqref{eqn:profit},
    \begin{multline*}
        E[R(p^*)] 
        = (p^*-r)d(p^*) - m\int^B_{C-d(p^*)}\left(d(p^*)-C+u\right)f(u)\ud u \\
        > (\bar{p}-r)d(\bar{p}) - m\int^B_{C-d(\bar{p})}\left(d(\bar{p})-C+u\right)f(u)\ud u \\
        > (\bar{p}-r)d(\bar{p}) - (d(\bar{p})-C+B)m\cdot\Pr\big(\epsilon>C-d(\bar{p})\big)
    \end{multline*}
    The first inequality is due to the optimality of $p^*$, and the second due to the fact that $d(\bar{p})-C+B\ge d(\bar{p})-C+u$.
    $p^* < \bar{p}$, thus $E'[R(\bar{p})]<0$ due to quasiconcavity. From \eqref{eqn:E'}
    \begin{equation*}
      m\cdot\Pr\big(C-d(\bar{p})\big) < \bar{p} + \frac{d(\bar{p})}{d'(\bar{p})} -r.
    \end{equation*}
    $\bar{p} + \frac{d(\bar{p})}{d'(\bar{p})} = \bar{r}$ from \eqref{eqn:pbar}. Thus,
    \begin{eqnarray*}
        E[R(p^*)] 
        &>& (\bar{p}-r)d(\bar{p}) - (d(\bar{p})-C + B)(\bar{r} - r) \\
        &=& (\bar{p}-\bar{r})d(\bar{p}) + (\bar{r}-r)(d(\bar{p})-d(\bar{p})+C-B) \\
        &=& E[\bar{R}(\bar{p})] + (\bar{r}-r)(C-B).
    \end{eqnarray*}
$C>B$ always holds as discussed in Sec.~\ref{subsec:profit_model}.